\documentclass[secnumarabic,amssymb, nobibnotes, nofootinbib,aps,pra]{revtex4-1}
\usepackage[dvipdfmx]{graphicx}
\usepackage{hyperref}
\usepackage{amsmath}
\usepackage{epsfig}
\usepackage{bm}
\usepackage[usenames]{color}

 \def\0ex{0_{\mathrm{ex}}}

\usepackage{braket} 
\usepackage{graphicx}
\usepackage{amsmath,amssymb}
\usepackage{bm}
\usepackage{color}
\usepackage{MnSymbol}

\def\>{{\rangle}}
\def\<{{\langle}}
\def\)>{{)\!\rangle}}
\def\(<{{\langle\!(}}
\def\C{\cal}
\def\B{\bar}
\graphicspath{{../figs/}}

\def\>{{\rangle}}
\def\<{{\langle}}

\begin{document}
   
\title{Dissipative dynamical Casimir effect in terms of 
the complex spectral analysis in the symplectic-Floquet space} 

\author{Satoshi \surname{Tanaka}}
\affiliation{Department of Physical Science, Osaka Prefecture 
University, Gakuen-cho 1-1, Sakai 599-8531, Japan}
\author{Kazuki \surname{Kanki}}
\affiliation{Department of Physical Science, Osaka Prefecture 
University, Gakuen-cho 1-1, Sakai 599-8531, Japan}

\small

\begin{abstract}

Dynamical Casimir effect of the optomechanical cavity interacting with one-dimensional photonic crystal is theoretically investigated in terms of the complex spectral analysis of Floquet-Liouvillian in the symplectic-Floquet space.
The quantum vacuum fluctuation of the intra-cavity mode is parametrically amplified by a periodic motion of the mirror boundary, and the amplified photons are spontaneously emitted to the photonic band.
We have derived the non-Hermitian effective Floquet-Liouvillian from the total system Liouvillian with the use of the Brillouin-Wigner-Feshbach projection method in the symplectic-Floquet space.
The microscopic dissipation process of the photon emission from the cavity has been taken into account by the energy-dependent self-energy.
We have obtained the discrete eigenmodes of the total system by non-perturbatively solving the nonlinear complex eigenvalue problem of the effective Floquet-Liouvillian, where the eigenmodes are represented by the multimode Bogoliubov transformation.
Based on the microscopic dynamics, the nonequilibrium stationary eigenmodes are identified as the eigenmodes with vanishing values of their imaginary parts due to the balance between the parametric amplification and dissipation effects.
We have found that the nonlocal stationary eigenmode appears when the mixing between the cavity mode and the photonic band is caused by the indirect virtual transition, where the external field frequency to cause the DCE can be largely reduced by using the finite bandwidth photonic band.

\end{abstract}

\date{\today}

\maketitle
\section{Introduction}

A vacuum fluctuation is one of the most characteristic features of quantum mechanics with no classical analog \cite{MilonniBook}.
Besides the well-known examples of the quantum vacuum fluctuation such as Lamb shift \cite{Lamb1947}, spontaneous emission \cite{Dirac1927,Weisskopf30ZeitPhys}, and static Casimir force\cite{Casimir48PR,Lamoreaux97PRL}, the dynamical Casimir effect (DCE) provides a direct method to observe the quantum vacuum fluctuation. 
The rapid motion of the boundary of an electromagnetic field invokes the conversion of localized virtual photons to real photons \cite{Moore70JMP,Fulling1976,Dodonov2010,Nation12RMP}.
The DCE has also attracted many researchers because of its close relation to Hawking radiation and the Unruh effect \cite{Nation12RMP}.
Even with this interest, it has been difficult to experimentally observe the DCE because we need to move a boundary almost at the same speed of light \cite{Moore70JMP,Fulling1976}.
The success of the observation of the DCE has been reported almost 40 years after the prediction by Moore by using a superconducting circuit to change the boundary condition of the optical transmission line, where they have revealed the quantum nature of the emitted field, such as entangled photons and the squeezing effect\cite{Wilson11Nature,Lahteenmaki2013a}.

 The DCE has also been regarded as a parametric amplification of a vacuum fluctuation\cite{Law94PRA,Dodonov2010}.
Parametric amplification is a well-known technique to amplify a weak signal to be observable by using a pumping external field \cite{landau1976mechanics,Husimi1953,Louisell1961,LoudonBook,walls2008quantum},
which is quantum mechanically attributed to the virtual transition interaction  yielding a squeezed vacuum state.
While recent studies have shown that the effect of the virtual transition interaction is manifested in an ultrastrong cavity QED \cite{Liberato07PRL,Liberato09PRA,Stassi13PRL,Qin18PRL,Ciuti05PRB}, the parametric resonance enhances the virtual transition interaction even for the small light-matter coupling.

Whereas the vacuum fluctuation is amplified inside the cavity, what we actually observe are real photons emitted to a free radiation field as a spontaneous emission process.
Therefore, it is important to clarify the microscopic mechanism of the  transition from an amplified virtual photon to a real photon emitted to free radiation field in order to understand the DCE.
Actually, the non-equilibrium stationary state generating a steady energy flow in the DCE is achieved as a result of the microscopic balance between the parametric amplification and the dissipation of the spontaneous emission.

With regards to theoretical analyses, the dissipation processes of the DCE have been described by the input-output theories \cite{Collett84PRA,Gardiner1985,Ciuti06PRA}, and the Lindblad type quantum master equations \cite{Carmichael87JOSAB,Kohler97PRE,Liberato09PRA}.
While these theories are mostly based on the Markov approximation justified for the emission to a free radiation field with infinite bandwidth, they are inappropriate to describe a spontaneous emission to a narrow-bandwidth photonic crystal with a bandgap \cite{John1990,John1994,DeLiberato2014,Calajo2017,Rybin2017}.
Recent advances in hybrid quantum systems, such as optomechanical systems where a photon emission process is manipulated at a single photon level, require a theory of DCE, taking into account a microscopic dissipation mechanism \cite{Xiang2013,Aspelmeyer2014,Settineri2019a}.

Recently, the microscopic descriptions of the dissipation process have been developed, known as complex spectral analysis \cite{Petrosky00PRA,Karpov2000b,Petrosky01PRA,Ordoez2001} and non-Hemitian quantum mechanics \cite{Hatano97PRB,Bender98PRL,MoiseyevBook,BenderPTsymmetry}.
In the complex spectral analysis, the functional space for a quantum state is extended to {\it  the rigged Hilbert space} where the dual functional space is equipped with a bi-complete and bi-orthonormal basis set\cite{Petrosky91Physica,Prigogine1992}, so that the time-evolution generator, Hamiltonian or Liouvillian, has  complex eigenvalues.
It has been clarified that when we change the system parameters, such as the discrete mode energy, to be in resonance with the continuum, there appears a bifurcation known as the {\it exceptional point} (EP) where the eigenvalues are changed to be from real  to complex.
We have applied this theory to open quantum systems to study dissipation processes of a discrete quantum state interacting with a continuum with a finite bandwidth.
We have revealed that the decay is nonanalytically enhanced when the discrete state is located closely to the bandedge of the continuum, and, as a result, it shows a nonanalytic decay process\cite{Tanaka06PRB,Tanaka2007,Tanaka13PRA,Tanaka16PRA}.
Therefore, in order to describe the DCE of a hybrid quantum system, it is important to study the competition between the parametric amplification and the   resonance instability which will be  clarified only when we  take into account the effect of the energy-dependent self-energy.

\begin{figure}
\begin{center}
\includegraphics[height=80mm,width=100mm]{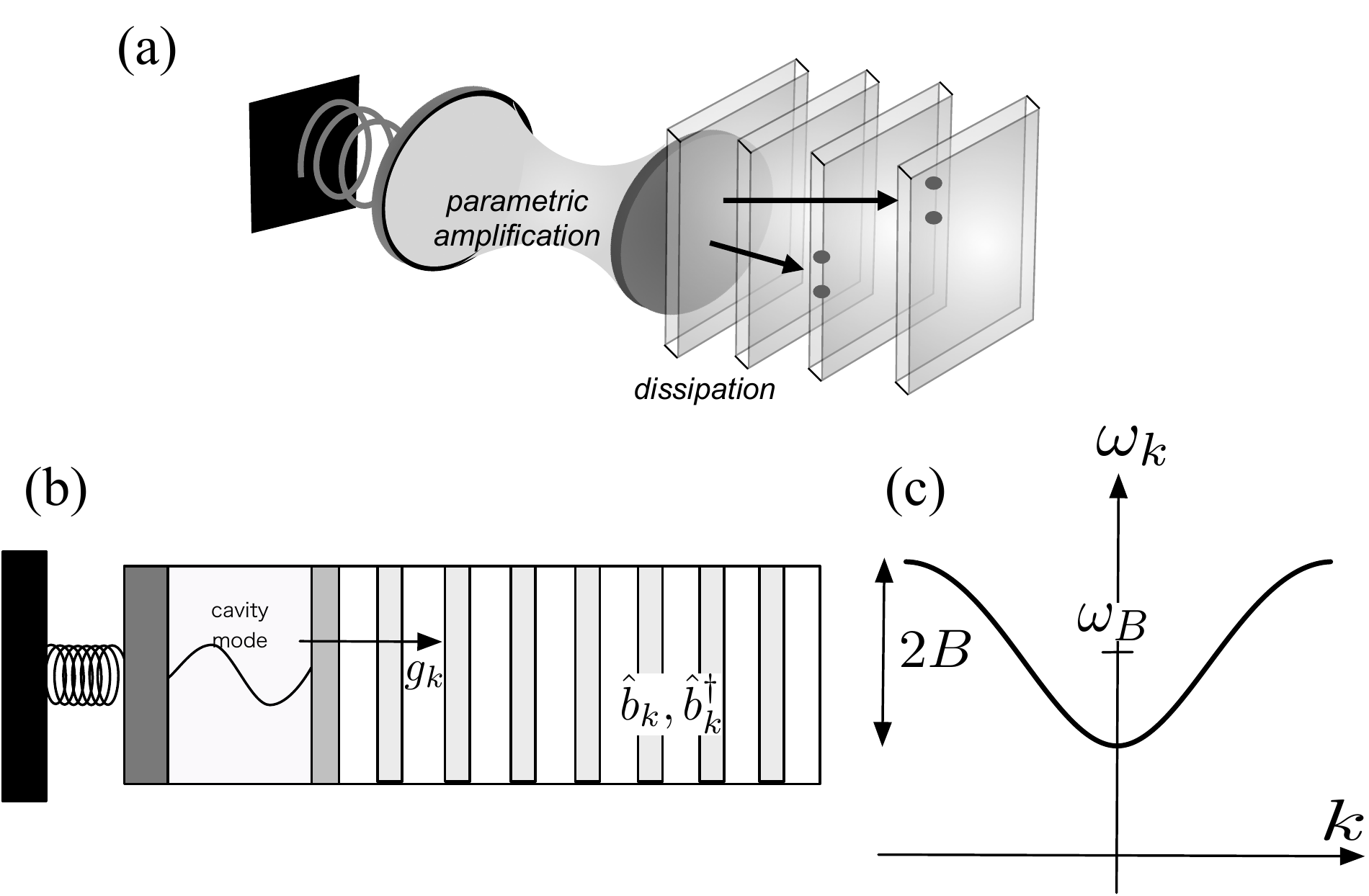}
\caption{(a) Optomechanical cavity interacting with a photonic crystal. (b) The frequency of a single cavity mode is periodically changed by mechanical pumping, and the cavity mode photon decays into one-dimensional photonic band. (c) Dispersion relation of one-dimensional photonic band with a bandwidth of $2B$ and the central frequncy $\omega_B$.}
\label{fig:hybrid}
\end{center}
\end{figure}

 In this paper, we theoretically study the parametric amplification of a quantum vacuum of an optomechanical cavity interacting with a photonic band, where the mirror boundary is periodically moved by a classical external force, as shown in Fig.\ref{fig:hybrid}(a)\cite{Settineri2018b}. 
The total system is composed of optomechanical cavity and photonic band states, and the time evolution of the canonical operators  obeys the Heisenberg equation, where the generator of the time evolution is determined by the commutator with the Hamiltonian, i.e. the Liouvillian superoperator\cite{Lowdin1985}.
While the Liouvillian is time-dependent and the system energy is no longer time-invariant due to the time-dependent external force, 
 the symplectic inner-product of the mode functions is time-invariant which ensures the existence of the canonical pair of the dynamical variables\cite{Moore70JMP,arnold2007mathematical,Landa_2012,meyer2013introduction}.

In this work, we study the time evolution of the dynamical variables as the symplectic transformation in the symplectic space.
With the use of the Floquet method\cite{Sambe73PRA,Kohler97PRE,Grifoni1998},  we have transformed the time-dependent problem to a time-independent eigenvalue problem to obtain  the eigenmodes of the total system in the symplectic-Floquet space\cite{RamirezBarrios2020}.
In the course of our analysis, the non-Hermitian effective Liouvillian is first derived in terms of the Brillouin-Wigner-Feshbach projection operator methods, where the microscopic dissipation process is rigorously taken into account with an energy-dependent self energy \cite{Feshbach62AnnalPhys,Rotter09JPhysA,Hatano2013,Kanki2017,Yamane18Symmetry}.
The complex eigenvalue problem of the effective Floquet-Liouvillian is solved to obtain new normal modes in terms of the multimode  Bogoliubov transformation, where the stationary mode is determined by the one with a vanishing imaginary part of its eigenvalue as a result of the balance between the parametric amplification and the dissipation.
We have found out the appearance of a non-local stationary modes as a result of the balance between the dissipation and the parametric amplification of the cavity mode and the photonic bands, when the cavity mode frequency lies in a photonic bandgap and that we can reduce the pump frequency to cause the DCE.

In Section \ref{Sec:Model}, we show the present model consisting of optomechanical cavity and photonic crystal, and the total Hamiltonian for the system.
The time evolution of the canonical variables are represented as the symplectic transformation in the symplectic space.
In Section \ref{Sec:Floquet}, with the use of the Floquet method, we transform the Heisenberg equation to the time-independent complex eigenvalue problem of Floquet-Liouvillian.
The effective Floquet-Liouvillian is derived by using Brillouin-Wigner-Feshbach projection method in Section \ref{Sec:Spectra}, where the microscopic dissipation effect is rigorously taken into account in terms of the energy-dependent self-energy.
The details of the derivation is shown in Appendix \ref{AppSec:FL}.
The self-consistent nonlinear complex eigenvalue problem of the effective-Liouvillian is numerically solved to obtain the complex eigenspectrum of the resonance modes, where the competition between the parametric amplification and the dissipation may be clear in comparison with a phenomenological calculation.
We shall reveal the effect of the finite width of the photonic band on the DCE that the nonlocal stationary mode appears as a result of the parametric mixing of the cavity mode and the photonic band.
The results are interpreted in terms of the perturbation analysis for the complex eigenvalue problem of the effective Floquet-Liouvillian, where the cancellation of the multimode parametric mixing and the dissipation effect will be clear.
The resonance modes of the total system are represented in terms of the multimode Bogoliubov transformation in Appendix \ref{AppSec:Modes}, where the distinction between the virtual photon and real photon components becomes clear as to whether the resonance effect is dominant.
We make concluding remarks in Section \ref{Sec:Conclusions}, where we explain the advantage of using the coupling with the photonic band as a possible method to reduce the external field frequency for the observation of the DCE.

%
%
%

\section{Model and Symplectic structure}\label{Sec:Model}

We consider  a hybrid quantum system consisting of optomechanical cavity and one-dimensional photonic crystal,  as shown in Fig.\ref{fig:hybrid}(a).
We assume a single mode to be allowed to exist in the cavity, and the one end of the boundaries is periodically moved by an external mechanical force with a frequency $\Omega$, resulting in the periodical change of the cavity mode frequency.
Through the other end of the cavity, the parametrically amplified photons in the cavity are emitted into the one-dimensional photonic band which is represented by  a semi-infinite tight-binding model as shown in Fig.\ref{fig:hybrid}(b).

In terms of a parametric amplification of quantum vacuum of  the single cavity mode, we may consider the effective Hamiltonian represented by \cite{Razavy1985PRD,Law94PRA}
\begin{align}
\label{Hcav}
\hat H(t)=\omega_0 \hat a^\dagger \hat a+f(t)(\hat a+\hat a^\dagger )^2 + \int \omega_k  \hat b_k^\dagger \hat b_k dk + \int g_k(\hat a^\dagger \hat b_k+\hat b_k^\dagger \hat a) dk \;,
\end{align}
where $\hat a^\dagger$ ($\hat a$) and $\hat b_k^\dagger$ ($\hat b_k$) are the creation (annihilation) operators of the cavity mode and the photonic band, respectively.
We take  $\hbar=1$ in the present paper. 
The second term of $\hat H(t)$ represent the periodical change of the cavity mode frequency by the external mechanical force with the oscillating amplitude   
\begin{align}
 f(t)=  f_0 \sin({\Omega t}+\theta)  \;,
 \end{align}
where $\Omega$, $f_0$, and $\theta$ are the pumping frequency, the amplitude, and the initial phase, respectively.
The  Hamiltonian (\ref{Hcav}) represents a quantum analog of the damped Mathieu equation\cite{Kohler97PRE,Landa_2012}.
Hereafter, we take the origin of time $t_0=-\theta/\Omega$, and redefine $t=t-t_0$.

The dispersion relation of the photonic band described by a semi-infinite one-dimensional tight-binding model is given by
\begin{align}
\omega_k=\omega_B-B\cos k \quad (0\leq k \leq \pi)\;,
\end{align}
as shown in Fig.\ref{fig:hybrid}(c).
The interaction of the cavity with the photonic band is described by the last term of (\ref{Hcav}), where the coupling strength for each $k$ mode is given by $g_k=g B \sin k/\sqrt \pi$ with a dimensionless coupling constant $g$.
We have adopted the rotating wave approximation as for the interaction between the cavity mode and the photonic band.
Note that we have taken the infinite limit of the total system's degrees of freedom $f$ resulting in the continuous wavenumber variables  $k$ in (\ref{Hcav}).

The time evolution of the Heisenberg operators is given by
\begin{align}\label{Heisen1}
-i{d\over dt}
 \begin{pmatrix}
\hat a \\
\hat b_k \\
\vdots\\
\hat a^\dagger \\
\hat b_k^\dagger\\
\vdots
\end{pmatrix} 
=
\begin{pmatrix}
-\omega_0 -2 f(t) & -g_k & \hdots & -2 f(t)  & 0 & \hdots\\
-g_k & -\omega_k            & \hdots &   0 & 0    & \hdots\\
\vdots& \vdots&\ddots&\vdots& \vdots & \ddots\\
2 f(t)  & 0       & \hdots  &  \omega_0+2 f(t)  & g_k  &\hdots\\
0 &  0 & \hdots & g_k &  \omega_k  &\hdots\\
\vdots& \vdots&\vdots&\vdots&\vdots & \ddots\\
\end{pmatrix}
 \begin{pmatrix}
\hat a \\
\hat b_k \\
\vdots\\
\hat a^\dagger \\
\hat b_k^\dagger\\
\vdots
\end{pmatrix} 
\equiv {\cal L}(t)  
\begin{pmatrix}
\hat a \\
\hat b_k \\
\vdots\\
\hat a^\dagger \\
\hat b_k^\dagger\\
\vdots
\end{pmatrix} \;,
\end{align}
where $2f \times 2f$ matrix ${\cal L}(t)$ is a representation matrix of the superoperator  $[\hat H(t),\cdot]$ which we shall call the Liouvillian.
We represent the solution of the Heisenberg equation (\ref{Heisen1}) by a $2f$-dimensional column vector of the operators as
\begin{align}\label{OperatorVector}
|\hat\Psi(t)\>_S= \left(\hat a(t), \{\hat b_k(t)\}, \hat a^\dagger(t), \{\hat b_k^\dagger(t)\}\right)^T \;,
  \end{align}
 where $T$ denotes the transpose and  $\{\cdots \}$ denotes all the wavenumbers of the photonic band.
 In (\ref{OperatorVector}) the first and the second half elements correspond to the annihilation and the creation operators, respectively, and  the meaning of the suffix  $S$ will be seen below.

 Introducing the basis set of $\left(|a\>_S,\{|b_k\>_S\}, |a^*\>_S,\{|b_k^*\>_S\}\right)$ represented by the column vectors 
\begin{align}\label{symplecticBasis}
&|a\>_S\equiv (1,0,\cdots ;0,0,\cdots)^T \;,\; |b_k\>_S \equiv (0,\cdots,1,\cdots ; 0,\cdots)^T \;, \notag\\
&|a^*\>_S\equiv (0,0,\cdots ;1,0,\cdots)^T \;,\; |b_k^*\>_S \equiv (0,\cdots,0,\cdots ; 0,\cdots ,1,\cdots)^T \;,
\end{align}
we express the solution of the Heisenberg equation (\ref{OperatorVector}) as 
\begin{align}\label{FieldOpBare}
|\hat\Psi(t)\>_S=\hat a(t) |a\>_S + \hat a^\dagger(t)  |a^*\>_S 
+ \int \left( \hat b_k(t)|b_k\>_S+\hat b_k^\dagger(t)|b_k^*\>_S \right)dk \;,
\end{align}
with the initial condition
\begin{align}\label{Psi0}
|\hat \Psi (0)\>_S= \hat a |a\>_S + \hat a^\dagger  |a^*\>_S 
+ \int \left( \hat b_k |b_k\>_S+\hat b_k^\dagger |b_k^*\>_S \right)dk.
\end{align}
The basis set of (\ref{symplecticBasis}) spans the $2f$-dimensional complex vector space, the {\it symplectic space} ($\C S$-space) \cite{Moore70JMP}.

When the time evolution of the field is represented by the $2f\times 2f$ time propagator matrix ${\C S}(t;t_0)$ as
\begin{align}\label{TimeEvolPsi}
|\hat\Psi(t)\>_S={\cal S}(t;t_0)|\hat\Psi(t_0)\>_S\;,
\end{align}
 ${\cal S}(t;t_0)$ obeys
\begin{align}\label{EQMofU}
-i{d\over dt}{\cal S}(t;t_0)={\cal L}(t){\cal S}(t;t_0)\;, 
\end{align}
with the initial condition
\begin{align}
{\cal S}(t_0;t_0)={\cal I}_{S} \;.
\end{align}
In this work, we take $t_0=0$, and cease to write $t_0$, and a curly character represents an operator (matrix) in the $\C S$-space.

Note that the time-evolution generator ${\C L}(t)$  possesses the symplectic  symmetry in the $\C S$-space
\begin{align}\label{SymplecticSymmetryL}
{\cal L}^T(t)={\cal J}{\cal L}(t){\cal J} \;,
\end{align}
where  the metric ${\C J}$ in the $\C S$-space is defined by
\begin{align}
{\cal J}\equiv \begin{pmatrix}
0 & I_f \\
-I_f & 0
\end{pmatrix} \;
\end{align}
with  an $f$-dimensional identity matrix $I_f$.
It  follows from (\ref{EQMofU}) and (\ref{SymplecticSymmetryL}) that ${\C S}(t)$ is symplectic 
\begin{align}\label{UtSymplec}
{\cal S}^T(t) {\cal J}{\cal S}(t)={\cal J} \;,
\end{align}
yielding the symplectic group Sp($2n,\mathbb{C}$)\cite{Moore70JMP,Dutta1995,meyer2013introduction}.
Therefore, the time evolution from $|\hat\Psi(0)\>_S$ to $|\hat\Psi(t)\>_S$ in (\ref{TimeEvolPsi}) is regarded as the symplectic transformation in the $\C S$-space.
With the use of (\ref{UtSymplec}), indeed, it can be shown that under the symplectic transformation by ${\C S}(t)$, the symplectic inner-product of the vectors in the $\C S$-space is time-invariant:
\begin{align}
 \{|f(t)\>,|g(t)\>\}_S=\{{\C S}(t)|f(0)\>,{\C S}(t) |g(0)\>\}_S =  \{|f(0)\>,|g(0)\>\}_S \;, \;\text{ for} \forall |f(0)\>_S, |g(0)\>_S \in{\C S}  
\end{align}
where  the {\it symplectic inner-product} is defined by 
\begin{align}
 \{|f\>,|g\>\}_S\equiv |f\>_S^T {\C J}|g\>_S \;.
\end{align}
Therefore, the symplectic inner-product of the basis set  (\ref{symplecticBasis}) is also time-invariant: 
\begin{align}
\{|j(t)\>,|j'(t)\>\}_S\equiv\{{\C S}(t)|j(0)\>,{\C S}(t)|j'(0)\>\}_S=\{|j\>,|j'\>\}_S=({\C J}_{j,j'})\;,\; (j,j'=a,\{b_k\},a^*,\{b_k^*\})\;.
\end{align}
Substituting (\ref{Psi0}) into (\ref{TimeEvolPsi}), we may alternatively represent $|\hat \Psi(t)\>$ defined by (\ref{FieldOpBare}) in terms of the time-evolved basis as
\begin{align}\label{hatPsit2}
|\hat \Psi (t)\>_S= \hat a |a(t)\>_S + \hat a^\dagger  |a^*(t)\>_S 
+ \int \left( \hat b_k |b_k(t)\>_S+\hat b_k^\dagger |b_k^*(t)\>_S \right)dk\;,
\end{align}
where $|j(t)\>_S\equiv {\C S}(t)|j\>_S$ $(j=a,\{b_k\},a^*,\{b_k^*\})$.

Associated with  the $\C S$-space, we define  the dual space, $\tilde {\C S}$-space, whose dual basis $(_{\tilde S}\<\tilde j|, {}_{\tilde{ S}}\<{\tilde j}^*|)$, is defined by 
\begin{align}\label{leftbasis}
_{\tilde S}\<\tilde a|\equiv |a\>^T \;,\; _{\tilde S}\<\tilde b_k| \equiv |b_k\>^T \;, 
\; _{\tilde S}\<\tilde a^*|\equiv |a^*\>^T \;,\; _{\tilde S}\<\tilde b_k^*| \equiv |b_k^*\>^T
\end{align}
so that the symmetric inner product with the basis of $\C S$-space is given by\cite{meyer2013introduction}
\begin{align}
_{\tilde S}\<\tilde j|j'  \>_S\equiv \delta_{j,j'} \;,\;   (j=a,\{b_k\},a^*,\{b_k^*\})\;.
\end{align}
The left basis set together with the right-basis defined in (\ref{symplecticBasis}) forms the bi-completeness relation in the $\tilde{\C S}\otimes {\C S}$-space:
\begin{align}\label{bicomplete}
{\C I}_{{\tilde S}S}=|a\>_{S}{}_{\tilde{S}}\< \tilde a|+ |a^*\>_ S{}_{\tilde{S}}\< \tilde a^*|
+  \int dk\left( |b_k\>_{ S}{}_{\tilde{ S}}\< \tilde b_k|+ |b_k^*\>_{ S}{}_{\tilde{ S}}\< \tilde b_k^*| \right)  \;.
\end{align}
Multiplying the left-basis (\ref{leftbasis}) from the left  to (\ref{hatPsit2}) and equating it with (\ref{FieldOpBare}), we have 
\begin{align}\label{BareSolutions}
\begin{pmatrix}
\hat j(t) \\
\hat j^\dagger(t)
\end{pmatrix}
=\begin{pmatrix}
\{ \<\tilde j|{\C S}(t) |j'\> \} & \{ \<\tilde j|{\C S}(t) |j'^*\> \} \\
\{ \<\tilde j^*|{\C S}(t) |j'\> \} & \{ \<\tilde j^*|{\C S}(t) |j'^*\> \}
\end{pmatrix}
 \begin{pmatrix}
\hat j' \\
\hat j'^\dagger
\end{pmatrix}
=\begin{pmatrix}
\{ \<\tilde j|j'(t)\> \} &  \{ \<\tilde j|j'^*(t)\> \} \\
\{ \<\tilde j^*|j'(t)\> \} &  \{ \<\tilde j^*|j'^*(t)\> \} 
\end{pmatrix}
 \begin{pmatrix}
\hat j' \\
\hat j'^\dagger
\end{pmatrix}
\;,\; (j= a, b_k)\;,
\end{align}
where  we have denoted the $2f$-dimensional column vector of the operators $(\hat a(t), \{ \hat b_k(t) \}, \hat a^\dagger(t), \{ \hat b_k^\dagger(t) \})^T$ by $(\hat j(t),\hat j^\dagger(t))^T$, and  have eliminated the suffix $S$ (or $\tilde S$)  of the vectors avoiding a heavy notation. 
The $2f\times 2f$ matrices in (\ref{BareSolutions}) are called the fundamental matrix solution, where  $ \{ \<\tilde j|{\C S}(t) |j'\> \}$ and $\{ \<\tilde j|j'(t)\>\}$ denote the $f\times f$ block matrices\cite{meyer2013introduction}.
It is shown in (\ref{BareSolutions}) that the operator solutions of the Heisenberg equation are given by the Bogoliubov transformation of the bare operators with the representation of the matrix ${\C S}(t)$  in terms of  the left- and right-basis of  $\{|j\>,|j^*\>\}$ and $\{\<\tilde j|,\{\<\tilde j^*|\}$.
With the use of the symplecticity of the matrix $S(t)$, it can be shown that the commutation relation 
\begin{align}
[\hat j(t),\hat j'^\dagger (t)]=\delta_{j,j'}\;,\; [\hat j(t),\hat j' (t)]= [\hat j^\dagger(t),\hat j'^\dagger (t)]=0\;, 
\end{align}
holds at an arbitrary time $t>0$.
Our purpose is to find the symplectic transformation to write ${\C S}(t)$ in a diagonal form, which determines the eigenmodes of the total system.

As an example,  we illustrate the case of a time-independent system, i.e. ${\cal L}(t)={\cal L}$, where we show that ${\cal S}(t)$ is represented by the eigenstates of ${\cal L}$ in the ${\C S}$-space.
It is found from the symplectic symmetry of ${\C L}$ (\ref{SymplecticSymmetryL}) that the eigenvalues of ${\C L}$ are obtained as a pair of the opposite signs, $z$ and $\B z\equiv-z$, which we assign to an annihilation and a creation eigenmodes when the real parts of the eigenvalues are negative and positive, respectively.
The right-eigenvalue problems for the annihilation and creation modes read
\begin{align}\label{rightEVIndep}
{\C L}|\phi_\xi\>=z_\xi|\phi_\xi\> \;,\; {\C L}|\B\phi_\xi\>=\B z_\xi|\B\phi_\xi\> \;,  \; (\xi=1,\cdots ,f)
\end{align}
where $\xi$ is the indices of specifying the eigenmodes.
With the use of the bi-completeness (\ref{bicomplete}), we can represent
\begin{subequations}
\begin{align}
|\phi_\xi\>&=\sumint_{j=a,b_k}|j\>\left( \<\tilde j|\phi_\xi\>+|j^*\>\<\tilde j^*|\phi_\xi\>\right) \;,\; \\
|\B\phi_\xi\>&=\sumint_{j=a,b_k}|j\>\left( \<\tilde j|\B\phi_\xi\>+|j^*\>\<\tilde j^*|\B\phi_\xi\> \right)  \;,\quad (\xi=1,\cdots f)\;.
\end{align}
\end{subequations}
Then we define the right-eigenmatrix represented by 
\begin{align}\label{varPhiDef}
\varPhi \equiv 
\begin{pmatrix}
 \{ \<\tilde j|\phi_\xi\>\} &  \{\<\tilde j|\bar\phi_\xi\>\} \\
 \{ \<\tilde j^*|\phi_\xi\>\} &  \{\<\tilde j^*|\bar\phi_\xi\>\} 
\end{pmatrix}
\;, \; (j=a,b_k;\xi=1,\cdots,f) \;,
\end{align}
where $ \{ \<\tilde j|\phi_\xi\>\},  \{\<\tilde j|\bar\phi_\xi\>\},  \{ \<\tilde j^*|\phi_\xi\>\}$, and  $ \{\<\tilde j^*|\bar\phi_\xi\>\} $ denote the $f\times f$ block matrices.
The right-eigenvalue problem is rewritten by
\begin{align}\label{REVPhi}
{\cal L}\varPhi =\varPhi {\C Z} \;,
\end{align}
where $ {\C Z} $ is  the diagonal eigenvalue matrix
\begin{align}\label{DiagZindepT}
 {\C Z}\equiv {\rm diag}(\{z_\xi\},\{\bar z_\xi\}) \;.
\end{align}
It follows from the symplectic symmetry of ${\C L}$ (\ref{SymplecticSymmetryL}) that the eigenmatrix is symplectic\cite{meyer2013introduction}:
\begin{align}
{\varPhi}^T {\cal J}\varPhi={\cal J}\;.
\end{align}

Similarly the left-eigenvalue problem reads
\begin{align}
\<\tilde \phi_\xi| {\C L}=z_\xi\<\tilde\phi_\xi| \;,\; \<\tilde {\B\phi}_\xi| {\C L}=\B z_\xi\<\tilde{\B \phi}_\xi| \;.
\end{align}
and the left-eigenmatrix is defined by
\begin{align}
\tilde\varPhi \equiv 
\begin{pmatrix}
 \{ \<\tilde\phi_\xi | j \>\} &  \{ \<\tilde\phi_\xi | j^* \>\} \\
 \{ \<\tilde{\B \phi}_\xi | j \>\} &  \{ \<\tilde{\B\phi}_\xi | j^* \>\}
\end{pmatrix}
\;, \; (j=a,b_k;\xi=1,\cdots,f) \;.
\end{align}
The left-eigenvalue problem of ${\cal L}$ is also written by
\begin{align}\label{LEVPhi}
 \tilde\varPhi  {\cal L}={\C Z}\tilde\varPhi \;,
\end{align}
where we see the symplecticity 
\begin{align}
\tilde{\varPhi} {\cal J}\tilde\varPhi^T={\cal J}  \;.
\end{align}
By comparison with (\ref{REVPhi}) and (\ref{LEVPhi}), we find
\begin{align}\label{tildePhiandPhi}
\tilde{\varPhi}={\C J}^T\varPhi^T {\C J}\;,\; \tilde{\varPhi}\varPhi={\C I}\;,
\end{align}
where we have used the relation
\begin{align}
{\C J}{\C Z}+{\C Z} {\C J}=0 \;.
\end{align}
The relations (\ref{tildePhiandPhi}) gives the explicit relations of the matrix elements between $\varPhi$ and $\tilde\varPhi$  as
\begin{align}\label{RelationLR}
\begin{cases}
\<\tilde\varphi_\xi|j\>=\<j^*|\bar\varphi_\xi\>\\
\<\tilde\varphi_\xi|j^*\>=-\<j|\bar\varphi_\xi\>
\end{cases}
\; , \quad
\begin{cases}
\<\tilde{\bar\varphi}_\xi|j\>=-\<j^*|\varphi_\xi\>\\
\<\tilde{\bar\varphi}_\xi|j^*\>=\<j|\varphi_\xi\>
\end{cases} \;.
\end{align}

With the use of these eigen-matrices,  the time propagator ${\C S}(t)$ is expressed by
\begin{align}
{\cal S}(t)=e^{i{\C L}t}=\varPhi \exp\left[ i {\C Z} t \right]\tilde \varPhi \;.
\end{align}
The eigenmode operators of the system $\{ \hat\psi_\xi,\hat{\bar \psi}_\xi \}$ are obtained in the form of the Bogolibov transformation:
\begin{align} \label{EigenOp}
\begin{pmatrix}
\hat\psi \\
\hat{\bar\psi}
\end{pmatrix}
=\tilde\varPhi
\begin{pmatrix}
 \hat j \\
\hat j^\dagger
\end{pmatrix} \;.
\end{align}

\section{Symplectic Floquet space}\label{Sec:Floquet}

Since the Liouvillian ${\C L}(t)$ is time periodic,  the Floquet method in the $\C S$-space can be used to obtain the eigenmodes of the present  system\cite{Sambe73PRA,Kohler97PRE,Grifoni1998,meyer2013introduction,Wiesel199481,RamirezBarrios2020}.
According to the Floquet-Lyapunov theorem, we write 
\begin{align}\label{RightFloquet}
{\C S}(t)= \varPhi(t) \exp[i {\C Z} t] \;,
\end{align}
with a Floquet periodic matrix 
\begin{align}
\varPhi (t+T)=\varPhi (t) \;,
\end{align}
where  $\varPhi(t)$ and ${\C Z}$ are represented in the same way as  (\ref{varPhiDef}) and (\ref{DiagZindepT}):
\begin{align}\label{varPhidef}
\varPhi (t) \equiv 
\begin{pmatrix}
 \{ \<\tilde j|\varphi_\xi(t)\>\} &  \{\<\tilde j|\bar\varphi_\xi(t)\>\} \\
 \{ \<\tilde j^*|\varphi_\xi(t)\>\} &  \{\<\tilde j^*|\bar\varphi_\xi(t)\>\} 
\end{pmatrix}\;,  \;
{\C Z}={\rm diag}(\{z_\xi\},\{\B z_\xi\}) \;.
\end{align}

Substituting (\ref{RightFloquet}) into (\ref{EQMofU}), we have derived the right-eigenvalue problem of the Floquet-Liouvillian ${\C L}_{\rm F}(t)$ in the $\C S$-space as
\begin{align}\label{FloquetEq}
{\cal L}_{\rm F}(t) \varPhi(t)\equiv \left[ {\cal L}(t)+i{d\over dt}\right] \varPhi(t)=\varPhi (t){\C Z}  \;.
\end{align}
Since ${\C S}(t)$ is symplectic, so is $\varPhi(t)$:
\begin{align}
\varPhi(t)^T {\C J}\varPhi(t)={\C J}\;.
\end{align}
%
When we define
\begin{align}\label{RelationLRPhi}
\tilde\varPhi (t)\equiv {\C J}^T \varPhi (t)^T {\C J} \;,
\end{align}
we find that $\tilde{\varPhi}(t)$ becomes the left-eigenmatrix of ${\C L}(t)$:
\begin{align}
\tilde\varPhi(t){\C L}_F(t)= \tilde\varPhi(t)  \left[ {\cal L}(t)+i\overleftarrow{{d\over dt}}\right] ={\C Z} \tilde\varPhi(t) \;.
\end{align}
with its matrix form as
\begin{align}
\tilde\varPhi (t)\equiv 
\begin{pmatrix}
 \{ \<\tilde\varphi_\xi (t) | j \>\} &  \{ \<\tilde\varphi_\xi (t)| j^* \>\} \\
 \{ \<\tilde{\B \varphi}_\xi(t) | j \>\} &  \{ \<\tilde{\B\varphi}_\xi (t) | j^* \>\}
\end{pmatrix}
\;.
\end{align}
It  follows from (\ref{RelationLRPhi}) that
\begin{align}
\tilde \varPhi (t)=\varPhi (t)^{-1} \;.
\end{align}
The relations of the matrix elements between $\varPhi(t)$ and $\tilde\varPhi(t)$ are given by (\ref{RelationLR}) .

The relations of $\varPhi(t)$ and $\tilde\varPhi(t)$ 
\begin{align}\label{orthonormalComplete}
\tilde\varPhi(t) \varPhi(t')={\cal I}\delta(t-t')\;,\; \varPhi(t)\tilde\varPhi(t')={\C I}\delta(t-t')\;,
\end{align}
represents the bi-orthonormality and bi-completeness in the $\C S$-space, respectively.
Indeed, the first equation reads
\begin{subequations}\label{BiComOrho}
\begin{align}
&\<\tilde\varphi_\xi(t)|\varphi_{\xi'}(t')\>=\<\tilde{\bar\varphi}_\xi(t)|\bar\varphi_{\xi'}(t')\>=\delta_{\xi,\xi'}\delta(t-t')\;,\; \\
 &\<\tilde\varphi_\xi(t)|\bar\varphi_{\xi'}(t')\>=\<\tilde{\bar\varphi}_\xi(t)|\varphi_{\xi'}(t')\>=0\;,
\end{align}
\end{subequations}
and the second equation reads
\begin{align}
{\C I}=\sumint d\xi \left( |\varphi_\xi(t)\>\<\tilde\varphi_\xi(t)|+ |\bar\varphi_\xi(t)\>\<\tilde{\bar\varphi}_\xi(t)| \right) \;.
\end{align}

Now we shall solve the eigenvalue problem of the Floquet-Liouvillian (\ref{FloquetEq}) to represent ${\C S}(t)$ in terms of the spectral decomposition in the composite symplectic-Floquet space ${\C F}\equiv {\C S}\otimes {\C T}$,
where the composite space ${\cal F}\equiv {\cal S}\otimes{\cal T}$ is composed of the symplectic vector  space  ${\cal S}$  and the space  ${\cal T}$  of periodic functions in time with period $T$ \cite{Grifoni1998}.
The details of how to construct the $\C F$-space is shown in Appendix  \ref{AppSec:FL}.
In the $\C F$-space, we transform the time-dependent differential equation (\ref{FloquetEq}) to time-independent eigenvalue problem of the Floquet-Liouvillian in terms of the Floquet mode representation\cite{Sambe73PRA,Kohler97PRE,Grifoni1998,RamirezBarrios2020}, as shown in (\ref{AppEq:LF}).
In the $\C F$-space, the symplectic Floquet eigenmatrix $\varPhi (t)$ in the $\C S$-space is expressed as a $t$-component of the symplectic  Floquet-vector $|\varPhi )$ by 
\begin{align}
\varPhi (t)\equiv (t|\Phi)=
\begin{pmatrix}
 \{ \<\!( \tilde j ,t |\varphi_\xi)\!\>\} &  \{\<\!( \tilde j,t|\bar\varphi_\xi)\!\>\} \\
 \{ \<\!( \tilde j^*,t|\varphi_\xi)\!\>\} &  \{\<\!( \tilde j^*,t|\bar\varphi_\xi)\!\>\} 
\end{pmatrix} \;,
\end{align}
where we have used (\ref{varPhidef}).


The  eigenvalue problem of the symplectic Floquet-Liouvillian in the $\C F$-space then reads
\begin{align}\label{EVLFLandR}
{\C L}_{\rm F}|\varPhi)=  |\varPhi){\C Z} \;,\; (\tilde\varPhi|{\C L}_{\rm F}={\C Z} (\tilde\varPhi| \;,
\end{align}
where the explicit matrix form of ${\C L}_{\rm F}$ in the $\C F$-space is given by (\ref{AppEq:LFfull})  in terms of the symplectic-Floquet mode basis set of  $\{ |j,\kappa_n)\!\>, | j^*,\kappa_n)\!\>\}$ $(j=a,b_k,\kappa_n=n\Omega; n=0,\pm 1,\cdots)$ .
Note that the Floquet-Liouvillian holds symplectic symmetry:
\begin{align}
{\C J}{\C L}_{\rm F}{\C J}={\C L}_{\rm F}^T \;.
\end{align}

\begin{figure}
\begin{center}
\includegraphics[height=80mm,width=100mm]{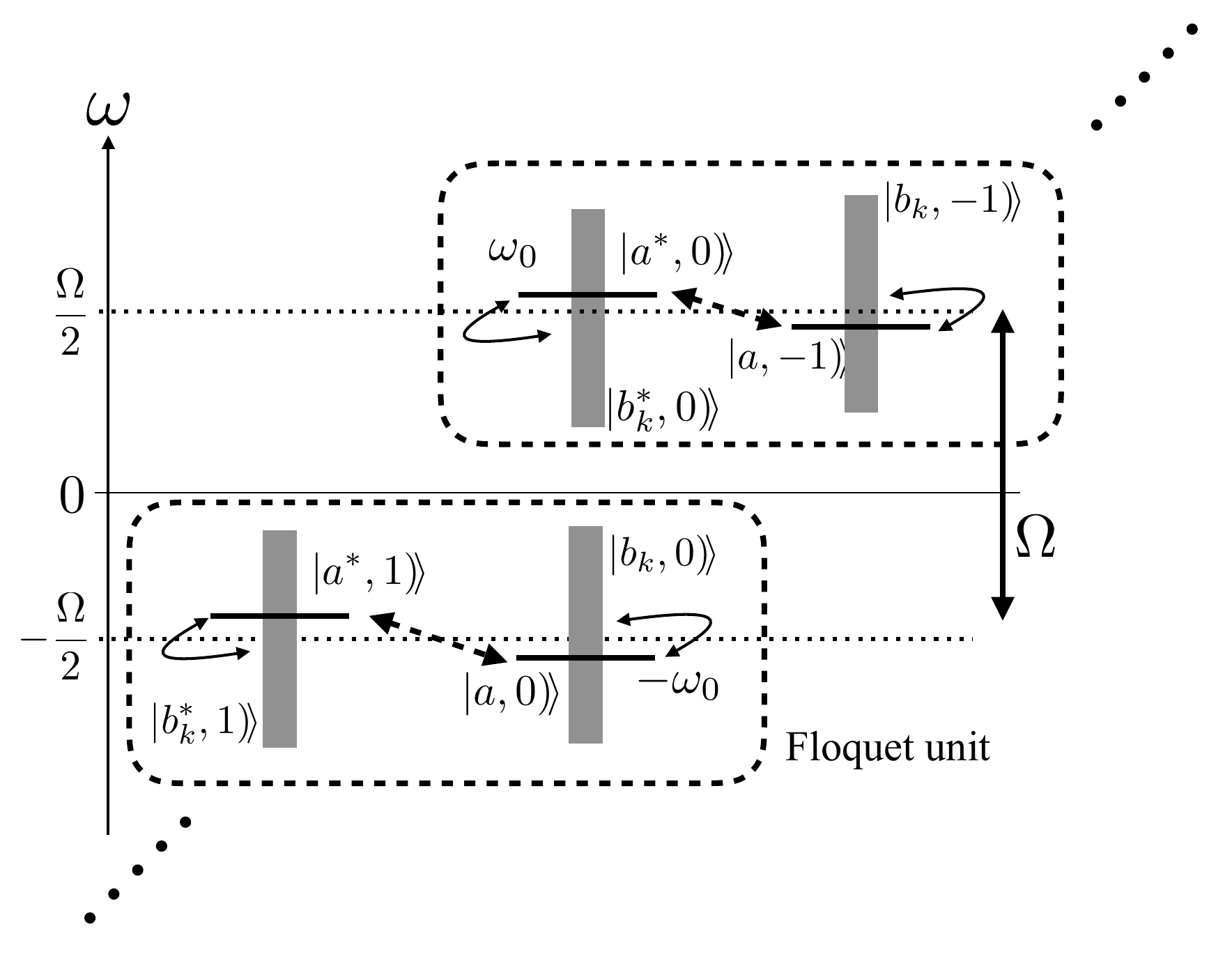}
\caption{Level scheme of the Floquet Liouvillian ${\C L}_{\rm F}$ in the $\C F$-space.
There are discrete levels $|a,n)\!\>$ ($|a^*,n)\!\>$) for each Floquet mode $\kappa_n$ which couples with the photonic band states $|b_k,n)\!\>$ ($|b_k^*,n)\!\>$) as indicated by the thin arrows. In addition, the virtual transition between $|a,n)\!\>$ and $|a^*,n)\!\>$ which causes parametric instability, indicated by the dashed arrows.
Two different types of the instability mechanisms coexist in the present system.
The dashed box indicates the Floquet unit that we focus on in the present analysis.
}
\label{fig:LevelFig}
\end{center}
\end{figure}

We have shown the level scheme of the Floquet-Liouvillian in the $\C F$-space in Fig.\ref{fig:LevelFig}.
Annihilation $|a,n)\!\>$  and creation modes $|a^*,n)\!\>$ of the Floquet cavity modes couple with the corresponding Floquet photonic band states $|b_k,n)\!\>$ and  $|b_k^*,n)\!\>$, respectively,  as indicated by the thin arrows.
When the cavity mode frequency $\omega_0$ is in resonance with the photonic band frequency, i.e. $|\omega_0-\omega_B|<B$, a cavity photon is spontaneously emitted out to the photonic band.
This is the first type of the instability, resonance instability.
 In addition, the annihilation and creation cavity modes of the adjacent Floquet modes, $|a,n)\!\>$ and $|a^*,n\pm 1)\!\>$, are coupled by the virtual transition which causes parametric instability, indicated by the dashed arrows.
This is the second type of instability.
In the present system, the balance between the resonance instability and parametric instability determines a non-equilibrium stationary photon emission in the DCE.
Since when the external field frequency $\Omega$ is close to twice the cavity mode frequency, i.e. $\Omega \simeq 2\omega_0$, the virtual transition causes efficiently the parametric amplification, we focus on the interactions between the states of the Floquet unit indicated by the dashed box in Fig.\ref{fig:LevelFig}.
In the next section, we construct the effective Floquet-Liouvillian of the cavity modes of the Floquet unit and illustrate the effect of the microscopic dissipation mechanism on the  cavity DCE.

\section{Complex spectra of the effective Floquet-Liouvillian, non-equilibrium stationary mode}\label{Sec:Spectra}

When $\Omega\simeq 2 \omega_0$, since parametric amplification is mainly attributed to the virtual interaction between $|a,n)\!\>$ and $|a^*,n+1)\!\>$, we may restrict ourselves to the subspace of $\left(| a,0)\!\>, \{ |b_k,0)\!\>\}, |a^*,1)\!\>, \{|b_k^*,1)\!\>\}\right)$.
The Floquet-Liouvillian matrix of the restricted subspace is represented by an infinite dimensional matrix:
\begin{align}\label{LF0}
{\cal L}_{\rm F}^{(0)}=
\begin{pmatrix}
-\omega_0+{\Omega\over 2}  & -g_k & -i f_0  &0    \\
  -g_k  & -\omega_k+{\Omega\over 2}  & 0 & 0  \\ 
     -if_0  & 0 & \omega_0-{\Omega\over 2}  & g_k    \\
 0&  0&   g_k &   \omega_k-{\Omega\over 2}  
\end{pmatrix}  \;,
\end{align}
where  we have written the matrix elements of a wavenumber $k$ which represents all the continuous photonic band states.
In (\ref{LF0}),  we have shifted the diagonal matrix element by $\Omega/2$ to keep the symplectic symmetry, so that 
the eigenvalue matrix ${\C Z}'$  are shifted by
\begin{align}
{\C Z}'={\C Z}+{\Omega\over 2} 
\end{align}
 in the complex eigenvalue problem (\ref{EVLFLandR}).
By using Feshbach-Brillouin-Wigner projection method\cite{Feshbach62AnnalPhys,Rotter09JPhysA,Hatano2013,Kanki2017,Yamane18Symmetry}, we  incorporate the interaction of the cavity modes with the photonic bands into the energy-dependent self-energy in terms of the projection operators 
\begin{align}\label{ProjectionOp}
{\cal P}_{a}=  |a,0)\!\>\<\!(\tilde a,0|+ |a^*,1)\!\>\<\!(\tilde a^*,1|  \;,\; {\cal Q}_a\equiv 1-{\cal P}_a\;,
\end{align}
as shown in Appendix \ref{AppSec:FL}.
The formal expression of the effective Floquet-Liouvillian
\begin{align}
{\cal L}_{\rm eff}({\C Z}')\equiv {\cal P}{\cal L}_{\rm F} {\cal P}+ {\cal P}{\cal L}_{\rm F} {\cal Q}{1\over {\C Z}'-{\cal Q}{\cal L}_{\rm F}{\cal Q}}  {\cal Q}{\cal L}_{\rm F} {\cal P} \;.
\end{align}
results in the two-by-two matrix  in the $\{|a,0)\!\>, |a^*,1)\!\>\}$-subspace given by
\begin{align}\label{Leff}
{\cal L}_{\rm eff}(z')=
\begin{pmatrix}
  {\Omega\over 2}-\omega_0-\sigma\left({\Omega\over 2}-z'\right) &  -if_0   \\
  - if_0   &-\left({\Omega\over 2}-\omega_0\right)+\sigma\left({\Omega\over 2}+z'\right) \\ 
\end{pmatrix} \;.
\end{align}
%
For the present one-dimensional photonic band is represented by the semi-infinite chain tight-binding model, the self-energy is analytically given by\cite{Tanaka2007,Fukuta17PRA}
\begin{align}
\sigma(z)\equiv {B^2\over \pi}\int_{0}^\pi {g^2\sin^2k\over z-\omega_k} dk =g^2  \left( z-\omega_B-\sqrt{(z-\omega_B)^2-B^2}\right) \;,
\end{align}
where we will take $B=1$ as an energy unit in the present work.
Taking the parameter values as
\begin{align}
\omega_0'\equiv \omega_0-{\Omega\over 2}\;,\; \omega_B'\equiv \omega_B-{\Omega\over 2}\;,\; 
\end{align}
the effective Floquet-Liouvillian is written as
\begin{align}\label{Leff2}
{\cal L}_{\rm eff}(z')=
\begin{pmatrix}
  -\omega_0'+\sigma(z'+\omega_B') &  -if_0   \\
  - if_0   &  \omega_0'+\sigma(z'-\omega_B') \\ 
\end{pmatrix} \;,
\end{align}
where we shall take the second Riemann sheet for the two-valued self energy function defined by the Cauchy integral, so that the creation mode decays in time under the resonance situation \cite{Prigogine1977,Petrosky91Physica,Prigogine1992,Prigogine1999528}.
In (\ref{Leff2}) we have defined the self energies of the two Riemann sheets as
\begin{align}
\sigma_{\rm I}(z)\equiv g^2\left(z-\sqrt{z^2-1}\right) \;,\; \sigma_{\rm II}(z)\equiv g^2\left(z+\sqrt{z^2-1}\right) \;.
\end{align}
It is  seen in (\ref{Leff2}) that the virtual transition in the off-diagonal elements and the complex self-energy in the diagonal elements represent the parametric instability and the resonance instability, respectively.
Therefore, the effective Liouvillian describes the exponential instabilities in the DCE from a unified point of view.

The right-eigenvalue problem of the effective Liouvillian reads
\begin{align}
{\C L}_{\rm eff}(z_\xi'){\C P}_a|\varphi_\xi)\!\>=z'_\xi {\C P}_a|\varphi_\xi)\!\> \;,\; 
{\C L}_{\rm eff}({\B z}_\xi'){\C P}_a|\B\varphi_\xi)\!\>=\B z'_\xi {\C P}_a|\B \varphi_\xi)\!\> \;,\; 
\end{align}
It should be noted that this eigenvalue problem is nonlinear in the sense that the effective Liouvilian itself depends its own eigenvalue, and that the eigenvalues of the effective Liouvillian coincides with the total system Liouvillian only when we take into account the energy-dependent self-energy, as  in our previous studies \cite{Petrosky91Physica,Tanaka06PRB,Kanki2010,Yamada12PRB,Tanaka13PRA,Fukuta17PRA,Yamane18Symmetry,Tanaka2020Physics}.

\begin{figure}
\begin{center}
\includegraphics[height=50mm,width=100mm]{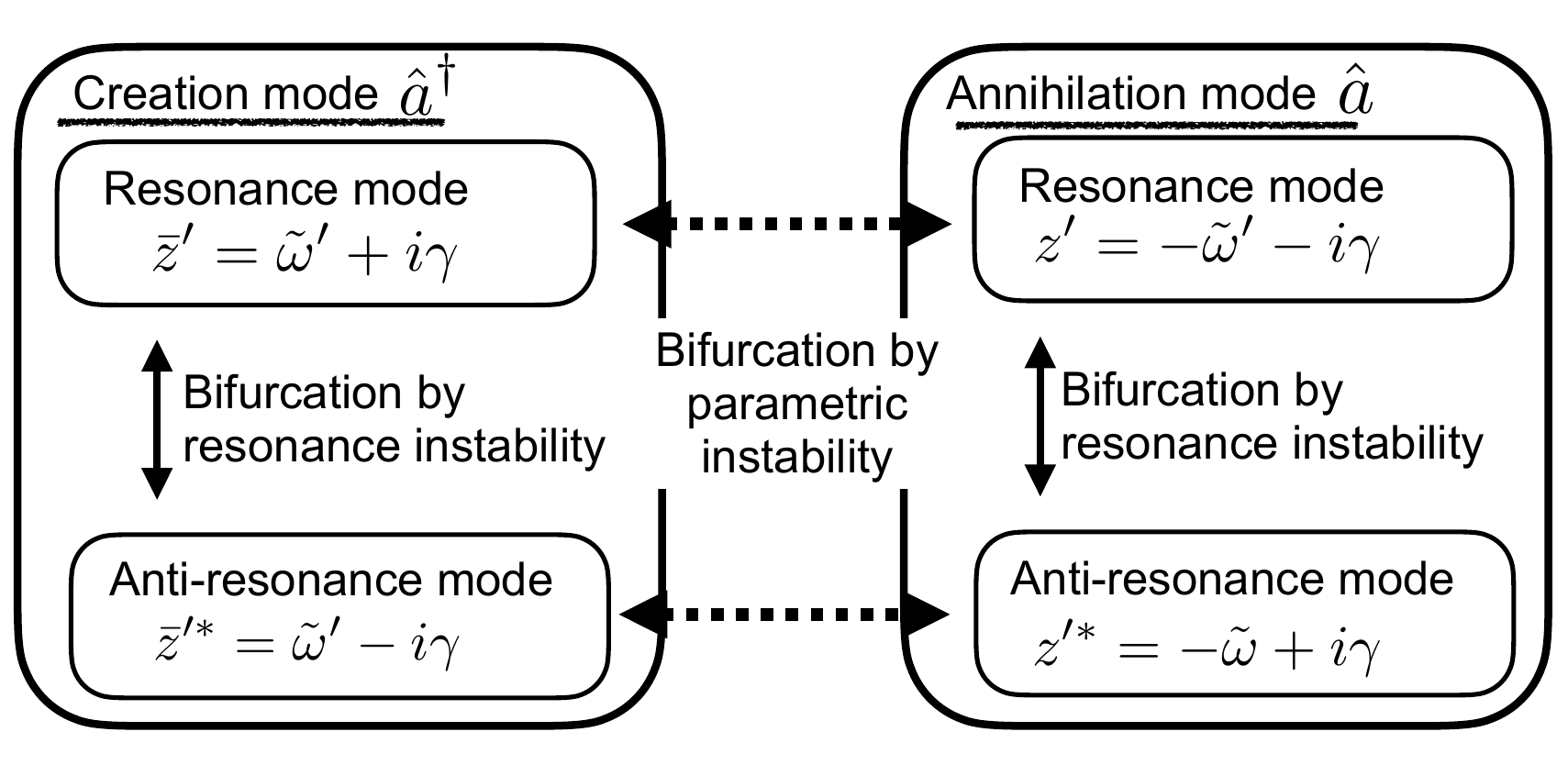}
\caption{Coupling scheme of the bifurcations of resonance and parametric amplification, where $\tilde\omega$ and $\gamma$ indicates the real part of the eigenvalues.}
\label{fig:Modes}
\end{center}
\end{figure}

The eigenvalues of (\ref{Leff}) are obtained by solving the characteristic equation
\begin{align}\label{dispersionEq}
\left\{z'+\omega_0'-\sigma(z'+\omega_B')\right\} \left\{z'-\omega_0'-\sigma(z'-\omega_B')\right\}  +f_0^2=0 \;.
\end{align}
It is seen that the equation remains the same by the change of $z\rightarrow \B z=-z$, by means of which we confirm that taking into account the energy dependence of the self-energy is essential to hold the  symplectic symmetry of the correct dynamics.
Correct consideration of the analytic continuation for the self-energy  brings about the four solutions.
The physical origin of the four solutions are assigned to a mixture of the resonance and antiresonance modes for each of cavity creation and annihilation modes, as shown in Fig.\ref{fig:Modes}.

\begin{figure}[t]
\centering
\includegraphics[height=70mm,width=120mm]{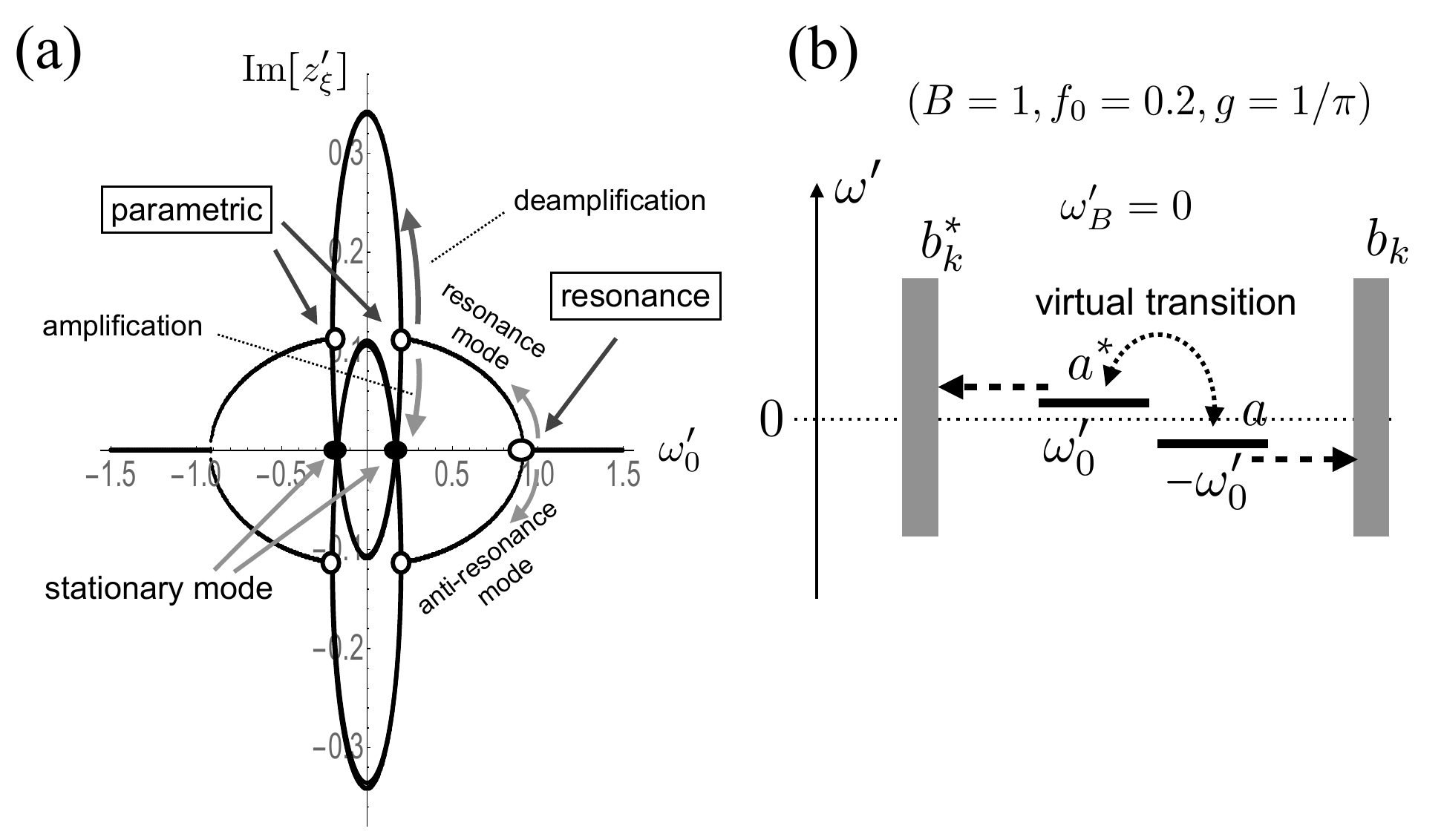}
\caption{(a) Imaginary part of the complex eigenvalues of ${\cal L}_{\rm eff}$ as a function of $\omega_0$ for the coupled optomechanical cavity with the photonic band, where the parameters are taken $\omega_B'=0 (\Omega=2\omega_B)$, $B=1$, $f_0=0.2$, and $g=1/\pi$.  
The horizontal axis represents $\omega_0' (\equiv\omega_0-\Omega/2)$. 
The bifurcation points are indicated by the open circles, and the stationary points are indicated by the black filled circles;
(b) Floque--Liouvillian level scheme of the $|a^*,1)\!\>, |a,0)\!\>, |b_k^*,1)\!\>$, and $|b_k,0)\!\>$ states, denoted by $a^*$, $a$, $b_k^*$, $b_k$ in the figure, respectively. 
The vertical axis denotes the frequencies of the modes in the Floquet--Liouvillian.
The dotted line is drawn at $-\Omega/2$ as a guide. }
\label{fig:Res1}

\end{figure}

In terms of the complex spectral analysis, we can identify the stationary mode whose eigenvalue has a vanishing value of the imaginary part, i.e., ${\rm Im}z_j=0$, as a result of the balance between the parametric amplification and the dissipation.
 We show the imaginary parts of the eigenvalues of ${\cal L}_{\rm eff}$ for $ \omega_B'=0\;(\Omega=2\omega_B )$ in Fig.\ref{fig:Res1}, where we change the cavity frequency $\omega_0$ while the values of $f_0=0.2$ and   $g=1/\pi$ are fixed.
In this case, the neighboring Floquet-photonic bands are overlapped,  as shown in Fig.\ref{fig:Res1}(b).

As $\omega_0'$ decreases, we encounter the bifurcation of the resonance instability at $\omega_0'=\omega_0-\Omega/2\simeq B= 1.0$, where the cavity mode becomes resonant with the photonic band, resulting in the bifurcation to resonance and anti-resonance modes.
In this figure, a~positive ${\rm Im} z'_\xi$ indicates a decaying direction as $t\to \infty$, seen from  (\ref{RightFloquet}).
With a further decrease of $\omega_0'$, the frequencies of the creation and annihilation cavity modes come close, and the effect of the virtual transition between them becomes significant.
Then, we encounter the second bifurcation of parametric instability at $\omega_0'\simeq f_0 =0.2$, where the downward and upward branches correspond to the parametric amplification and deamplification, respectively.
As we further decrease $\omega_0$, we reach the stationary point where ${\rm Im}z'_\xi=0$,  as a result of a balance between the parametric amplification and the dissipation effects, as indicated by the black filled circles.
At this point, the stationary energy flow coming out from the cavity to the photonic band is achieved with the spontaneous photon emission.
The figure clearly demonstrates that this stationary DCE has been determined by solving the dispersion equation (\ref{dispersionEq}) taking into account the energy dependence of the self-energy.

\begin{figure}
\centering
\includegraphics[height=60mm,width=90mm]{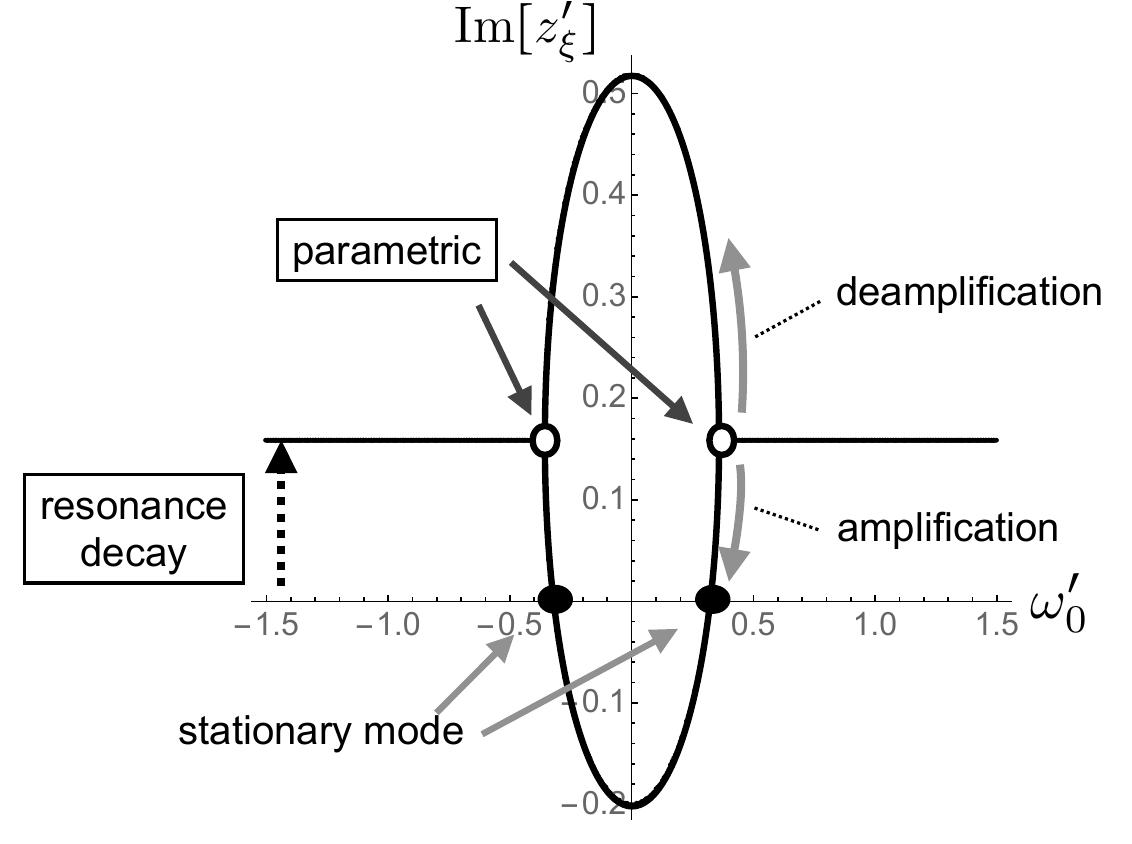}
\caption{Imaginary part of the complex eigenvalues of the phenomenological model  as a function of $\omega_0$ for the values of $f_0=0.2$ and $\gamma=1/\pi$, where the horizontal axis represents $\omega_0-\Omega/2$.}
\label{fig:ResPh}

\end{figure}

Here, we compare the present results with a phenomenological model of a damped parametric amplifier whose classical motion is represented by a damped Mathieu equation\cite{Kohler97PRE}
\begin{align} 
 \ddot{x}+\gamma \dot{x}+\omega(t)^2 x=0\;,
 \end{align}
where 
\begin{align}
 \omega^2 (t) =\omega_0^2\left( 1+ {4f_0\over \omega_0}\sin ( \Omega t ) \right) \;,
 \end{align}
and $\gamma$ is a phenomenological dissipation constant.
Corresponding Heisenberg equation reads
\begin{align}\label{DampedMathieu}
-i{d\over dt}
\begin{pmatrix}
\hat a \\
\hat a^\dagger 
\end{pmatrix} 
&=
\begin{pmatrix}
-\omega_0 - 2f(t) +i{\gamma\over 2} & - 2f(t) -i{\gamma\over 2}\\
 2f(t) -i{\gamma\over 2} &   \omega_0+ 2 f(t)  +i{\gamma\over 2}
\end{pmatrix}
 \begin{pmatrix}
\hat a \\
\hat a^\dagger 
\end{pmatrix} \;,
\end{align}
where the cavity mode operators are defined in a usual manner 
\begin{align}
\hat a={1\over\sqrt{2\omega_0}}\left(\omega_0 \hat x+i\hat p\right) \;,\;\hat a^\dagger={1\over\sqrt{2\omega_0}}\left(\omega_0 \hat x-i\hat p\right)  \;.
\end{align}
Similarly as above, by using Floquet-Lyapnov theorem, we can write down the Floquet-Liouvillian in the $\{|a,0)\!\>, |a^*,-1)\!\> \}$ subspace under the effective parametric amplification condition $\Omega\simeq 2\omega_0$.
The phenomenological Floquet--Liouvillian is then written as a constant matrix:
\begin{align}
{\cal L}_{\rm ph}=\begin{pmatrix}
-\omega_0'+i{\gamma\over 2}  & -if_0 -i{\gamma\over 2}  \\
-if_0 -i{\gamma\over 2}  &   \omega_0'+i{\gamma\over 2} 
\end{pmatrix} ,
\end{align}
where we shift the frequency by $\Omega/2$ as before.
The complex eigenvalue is immediately obtained by
\begin{align}\label{Phenomzj}
z'_\mp=z_\mp+{\Omega\over 2}= i {\gamma \over 2}  \mp i \sqrt{ \left( f_0 +{\gamma\over 2}\right)^2- \omega_0'^2 }\;.
\end{align}
In Fig.\ref{fig:ResPh},  we show the imaginary part of the solutions.
Within the parameter range of  $|\omega_0'|<f_0+\gamma/2$ the parametric amplification of the cavity mode happens and the stationary mode appears at 
\begin{align}
\omega_0'=\pm \sqrt{f_0(f_0+\gamma)}
\end{align}
as a balance between the parametric amplification and dissipation.
Meanwhile, since this phenomenological model assumes  a flat-band radiation with an infinite bandwidth, the resonance bifurcation does not appear.

\begin{figure}[t]
\centering
\includegraphics[height=80mm,width=130mm]{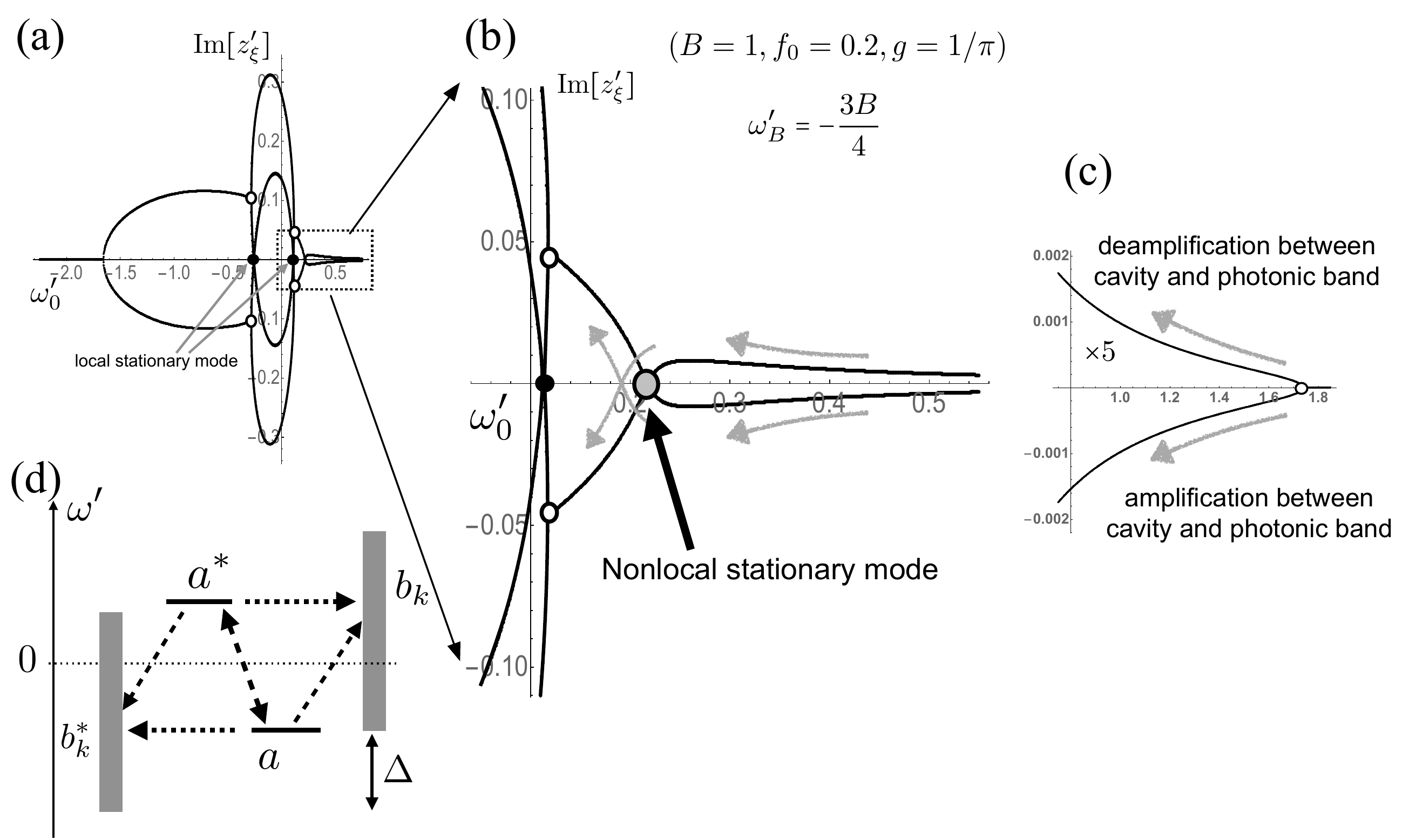}
\caption{(a)  Overall picture of the imaginary part of the complex eigenvalues of ${\cal L}_{\rm eff}$ as a function of $\omega_0$ for the coupled optomechanical cavity with the photonic band, where the parameters are taken $\omega_B'=-\Delta/2$, $\Delta=3B/2$, $B=1$, $f_0=0.2$, and $g=1/\pi$. 
The horizontal axis represents $\omega_0'$. 
The parametric bifurcation points are indicated by the open circles, and the stationary points are indicated by the filled circles. 
(b,c) Expanded pictures of the dotted box area in (a). 
Multimode stationary point is indicated by the gray filled circle. 
(d) Floquet--Liouvillian level scheme of the $|a^*,1)\!\>, |a,0)\!\>, |b_k^*,1)\!\>$, and $|b_k,0)\!\>$ states, where the vertical axis denotes the frequencies of the modes in the Floquet--Liouvillian.
}
\label{fig:Res2}
\end{figure}

The band edge effect is  pronounced when the two photonic bands for the creation and annihilation modes of the neighboring Floquet modes are shifted.
In Fig.\ref{fig:Res2}, we show the results  for $\omega_B'=-\Delta/2$ with $\Delta=3B/2$ so that the neighboring Floquet photonic bands are shifted by $\Delta$, as shown in Fig.\ref{fig:Res2}(d), where the other parameters are fixed at the same values of Figure \ref{fig:Res1}.
The overall behavior of ${\rm Im}z'_\xi$ is shown in Fig.\ref{fig:Res2}(a), where we have seen again the parametric bifurcation of the cavity mode indicated by the four open circles, and  the stationary points indicated by the two black filled  circles as a result of the balance between the resonance instability and the parametric amplification of the cavity modes. 

Nonetheless,  we find very different behaviors of the spectrum in the region of   $ \omega'_0>0.1B$ as indicated by the broken box in Fig.\ref{fig:Res2}(a), which is expanded in Figs.\ref{fig:Res2}(b) and (c).
For $\omega'_0\simeq -\omega'_B+B\lesssim 1.75 B$, the creation (annihilation) mode is in resonance with the annihilation (creation) photonic band modes, as shown in Fig.\ref{fig:Res2}(d).
Even though there is no direct virtual transition coupling between the cavity  and the photonic band modes, these modes can be indirectly coupled through the virtual coupling of the cavity modes.
Consequently, the multimode parametric amplification happens between the cavity and photonic bands at $\omega_0'\simeq 1.75 B$ as shown in Fig.\ref{fig:Res2}(c), where the multimode parametric bifurcation point is indicated by the open circle.
As $\omega'_0$ decreases, each cavity modes becomes in resonance with the corresponding photonic bands.
As a result, the decay into the photonic band and the multimode parametric amplification is balanced to yield a new type of the nonequilibtirum stationary mode around $\omega_0'\simeq \omega'_B+B= 0.25 B$ as indicated by the gray filled circle in Fig.\ref{fig:Res2}(b).
This nonequilibrium stationary mode is {\it nonlocal} in the sense that it is represented by a mixture of the cavity mode and the phtonic band modes.  Indeed, the eigenmode is obtained by a {\it multimode-Bogoliubov transformation} of the cavity mode and the photonic band of the total system as shown in Appendix \ref{AppSec:Modes} \cite{LoudonBook}.
As $\omega'_0$ further decreases, the direct decay to the photonic band becomes more effective, and the parametric amplification of the intra-cavity modes gives rise to the parametric bifurcation as shown by the open circles in Fig.\ref{fig:Res2}(b).

%

In order to understand the appearance of  the  nonlocal stationary mode as shown in Fig.\ref{fig:Res2}(b), we perturbatively consider the complex eigenvalue problem of the effective Floquet-Liouvillian (\ref{Leff2}).
Here we specifically consider the creation mode in ${\C L}_{\rm eff}(z')$.
The perturbed eigenvalue is obtained up to the second order of the virtual transition coupling $f_0$ as
\begin{subequations}\label{z2}
\begin{align}
\B z'^{(2)}&=\omega_0'+\sigma_{\rm II}(\omega_0'-\omega'_B)
-{f_0^2 \over 2\omega_0' +\sigma_{\rm II}(\omega'_0-\omega'_B)-\sigma_{\rm II}(\omega'_0+\omega'_B) } \\
&=\omega_0'+\sigma_{\rm II}(\omega_0'-\omega'_B) \notag\\
&-{f_0^2 \over R^2(\omega'_0)}\left\{ 2\omega'_0+{\rm Re}\left[\sigma_{\rm II}(\omega'_0-\omega'_B)-\sigma_{\rm II}(\omega'_0+\omega'_B)\right]
-i  {\rm Im}\left[\sigma_{\rm II}(\omega'_0-\omega'_B)-\sigma_{\rm II}(\omega'_0+\omega'_B)\right] \right\} 
\end{align}
\end{subequations}
where we denote
\begin{align}
R(\omega'_0)\equiv  \left( 2\omega_0' +{\rm Re}\left[\sigma_{\rm II}(\omega'_0-\omega'_B)-\sigma_{\rm II}(\omega'_0+\omega'_B)\right] \right)^2 
+\left( {\rm Im}\left[\sigma_{\rm II}(\omega'_0-\omega'_B)-\sigma_{\rm II}(\omega'_0+\omega'_B)\right]\right) ^2  \;.
\end{align}
In (\ref{z2}), 
the imaginary part is given by
\begin{align}
{\rm Im}\B z'^{(2)}&= {\rm Im} \sigma_{\rm II}(\omega'_0-\omega'_B)+
{f_0^2 \over R^2(\omega'_0)} {\rm Im}\left[\sigma_{\rm II}(\omega'_0-\omega'_B)-\sigma_{\rm II}(\omega'_0+\omega'_B)\right]  \notag \\
&=\left( 1+ {  f_0^2 \over R^2(\omega'_0)} \right) {\rm Im}[\sigma_{\rm II}(\omega'_0-\omega'_B)]
- { f_0^2\over R^2(\omega'_0)}{\rm Im}[\sigma_{\rm II}(\omega'_0+\omega'_B)] \;.
\end{align}
The first  term is attributed to the ordinary dissipation of the cavity creation mode to the  photonic creation band, where the analytic continuation of the self energy is taken such that the creation mode decays in time as mentioned in (\ref{Leff2}): ${\rm Im}\;\sigma_{\rm II}(\omega'_0-\omega'_B)>0$  for  $|\omega'_0-\omega'_B|<B$.
The second term is attributed to the indirect coupling  of the  cavity creation mode with the  photonic annihilation band via the virtual transition of the cavity modes, which gives the negative contribution to ${\rm Im} \B z'^{(2)}$.
This term causes the multimode parametric amplification between the cavity mode and the photonic band for $|\omega'_0+\omega'_B|<B$.
Therefore, when both resonance conditions of $\sigma_{\rm II}(\omega'_0-\omega'_B)$ and $\sigma_{\rm II}(\omega'_0+\omega'_B)$ are satisfied for $\Delta/2-B<\omega'_0<B-\Delta/2$,  where $\Delta\equiv -2\omega'_B$, the dissipation and the multimode parametric amplification is balanced to gives rise to the nonlocal stationary mode.
For $B-\Delta/2<\omega'_0$, the resonance condition for $\sigma_{\rm II}(\omega'_0-\omega'_B)$ is no longer satisfied, so that the multimode parametric amplification instability happens in the parameter rage of $B-\Delta/2<\omega'_0<B+\Delta/2$  with the multimode parametric bifurcation as shown in Fig.\ref{fig:Res2}(c).

%

We have obtained the explicit expression of the resonance eigenmodes in Appendix \ref{AppSec:Modes} where the balance between the parametric amplification and the resonance decaying is also well reflected in the form of the eigenmodes.
Although this simple perturbation analysis qualitatively  explains the cause of the nonlocal stationary mode, it is necessary to employ the non-perturbative analysis to determine the nonlocal stationary mode as shown in the present work.
The non-perturbative analysis  is especially required when the photonic band has a singularity in the density of states, such as Van Hove singularity in the one-dimensional  system \cite{Tanaka06PRB,Tanaka16PRA}.


\section{Conclusions}\label{Sec:Conclusions}

In this paper, we have studied the DCE of the optomechanical cavity interacting with a one-dimensional photonic crystal in terms of the complex spectral analysis of Floquet-Liouvillian in the symplectic-Floquet space, where the quantum vacuum fluctuation of the intra-cavity mode is parametrically amplified by a periodic motion of the mirror boundary, and the amplified photons are spontaneously emitted to the photonic band.
The virtual transition interaction of the cavity mode is enhanced by the parametric resonance with the external oscillating field. 
The effective non-Hermitian Floquet-Liouvillian has been derived from a Heisenberg equation of the total system by  using the Floquet method and the Brillouin-Wigner-Feshbach projection method, where we have taken into account a microscopic dissipation process in terms of the energy-dependent self energy. 
The non-Hermitian effective Floquet-Liouvillian has clarified the competing roles of the parametric amplification due to the virtual transitions and the dissipation due to the resonance.
The nonequilibrium stationary modes have been obtained as a result of the balance between the two instabilities, where the eigenmode of the Liouvillian is represented by the Bogoliubov transformation.
The photonic band edge effect is prominent when the cavity mode frequency is close to the band edge.
In this case, the indirect coupling between the cavity mode and photonic band via the virtual transition of the cavity modes yields the nonlocal stationary mode, which is represented by the multimode-Bogoliubov transformation of  the cavity mode and photonic band.

\begin{figure}
\centering
\includegraphics[height=70mm,width=100mm]{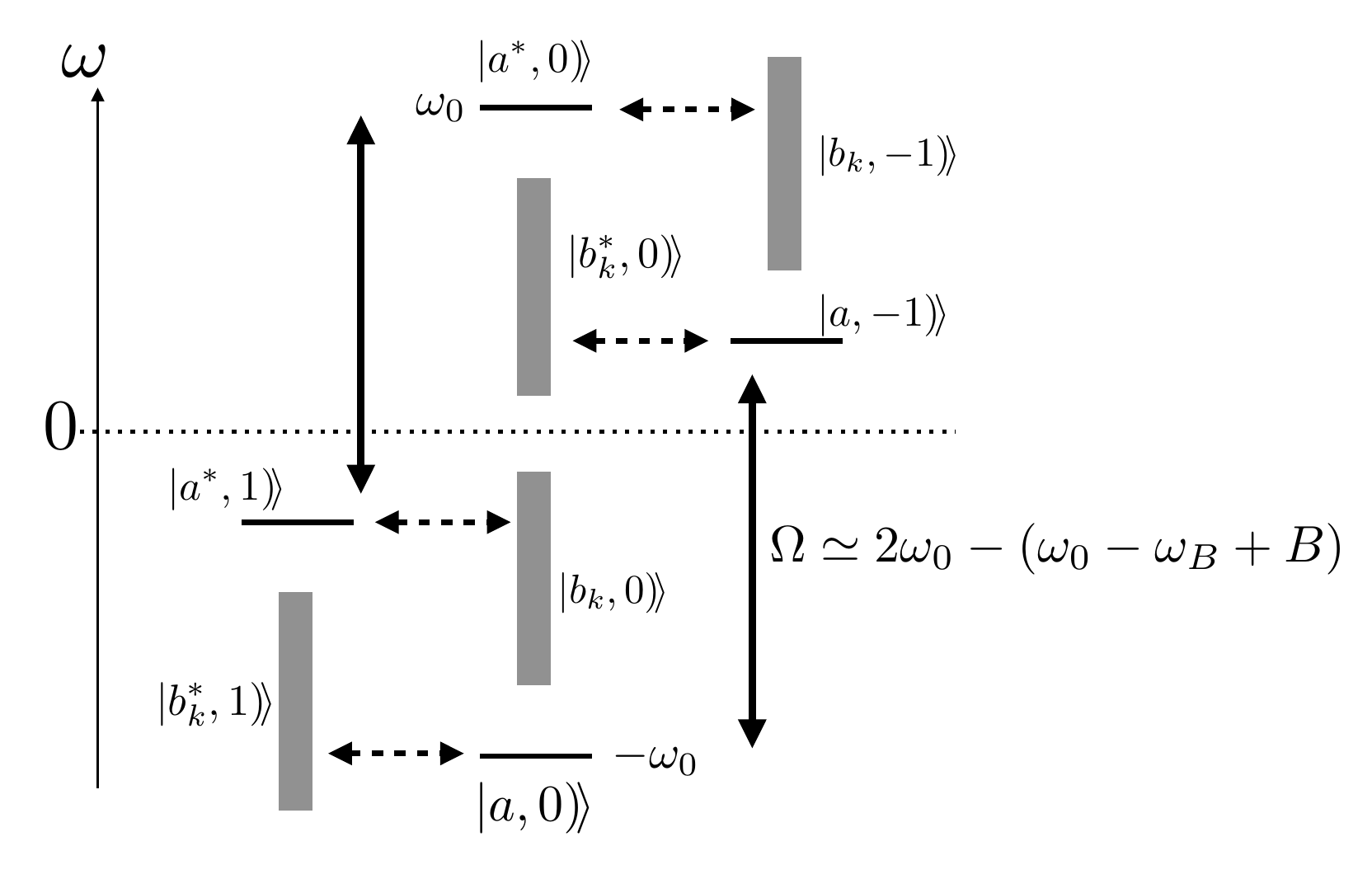}
\caption{ Floquet--Liouvillian level scheme in a situation direct dissipation to the photonic band is suppressed and multimode parametric amlification is induced with the external driving frequency $\Omega$ much smaller than $2\omega_0$.}
\label{fig:Reduction}

\end{figure}
Lastly, we would emphasize a practical advantage of the present model for the observation of the DCE.
A major obstacle for the observation of the DCE is the difficulty to move the boundary with almost twice the cavity frequency, $\Omega\simeq 2\omega_0$.
However, the results of the preceding section indicate that we may use the lower frequency pump field to induce the multimode parametric amplification, which is attributed to the indirect virtual coupling between the cavity mode and the photonic band while suppressing the direct dissipation to the photonic band.
In Figure \ref{fig:Reduction}, we show the frequency level scheme of $\hat a^\dagger$, $\hat a$, $\hat b_k^\dagger$, and $\hat b_k$ for that situation.
First,  in order to suppress the direct dissipation to the photonic band, the cavity mode frequency must be outside of the photonic band:
\begin{align}\label{cond1}
\omega_0>\omega_B+B.  \;
\end{align}

Under this condition, when the pumping field is taken as 
\begin{align}\label{cond2}
\Omega\sim 2\omega_0-(\omega_0-\omega_B+B) \;,
\end{align}
the creation (annihilation) cavity mode is in resonance with the annihilation (creation) mode of the photonic band so that the multimode parametric amplification happens.
It is clear from (\ref{cond1}) and~(\ref{cond2}) that the multimode parametric amplification happens for $\Omega < 2\omega_0-2B$, far smaller than $2\omega_0$.
Therefore, the major obstacle can be diminished.

Very recently, a method to reduce the mirror frequency for the DCE has been proposed to take advantage of a nonlinear interaction of the quantized mirror motion and the cavity photon mode~\cite{Settineri2019a,Macri2018}, where the strong nonlinear mixing between the mirror motion and cavity mode is assumed to be represented by the dressed state representation.
Our proposal provides an alternative method for the reduction of the pump frequency to induce the multimode parametric amplification by using a~finite bandwidth photonic band, i.e., a control of the dissipation process.
It is considered that the characteristics of the emitted photon are varied according to different types of the DCE emission~processes.

In the present work, we have found three different types of the stationary modes.
One is a stationary state well below the two bifurcation thresholds, where a cavity squeezed vacuum state is associated with a localized virtual photon cloud of the photonic band, and the periodic Rabi oscillation happens between the cavity squeezed vacuum and the virtual photon cloud.
The second one is the multimode DCE, where the stationary spontaneous photon emission to the photonic band happens with the two-photon entanglement between the cavity mode and the photonic band.
The third one is the ordinary DCE, where an entangled cavity photon pair is emitted to the photonic band.
We can observe the two-photon entanglement of the emitted photons by a quantum correlation observation, such as the homodyne detection method \cite{LoudonBook,walls2008quantum}.
The study of the real-time dynamics of these photon emission processes is now underway.

\section*{Acknowledgment}

We are very grateful T. Petrosky, R. Passante, H. Yamane, Y. Kayanuma, K. Mizoguchi, K. Noba, S. Garmon, and M. Domina for fruitful discussions.
This research was funded  by JSPS KAKENHI grants number  JP18K03496, JP17K05585, JP16H04003, and  JP16K05481.


%



%
%

%
%
%
%
%

\appendix

\section{Appendix A: Floquet-Liouvillian complex eigenvalue problem and effective operator}\label{AppSec:FL}

In this section, we briefly review the Floquet method according to Ref.\cite{Shirley1965,Sambe73PRA,Grifoni1998}, and derive the effective Floquet-Liouvillian in the symplectic space given in (\ref{Leff}).
The composite space ${\cal F}\equiv {\cal S}\otimes{\cal T}$ is composed of the symplectic vector  space  ${\cal S}$  and the space  ${\cal T}$  of periodic functions in time with period $T$ \cite{Grifoni1998}.
In ${\cal T}$-space, any periodic function of $t$  is represented as a vector
$f(t)\equiv (t|f)$, 
where the time basis is an eigenstate of a time operator
$\hat t|t)=t|t) $,
and the conjugate operator is given by
$\hat p_t\equiv i {\partial/\partial t}$ \;.
The eigenstate of $\hat p_t$ is given by
\begin{align}\label{AppEq:kappan}
 |\kappa_n)={1\over T}\int_0^T e^{i\kappa_n t}|t) dt  \;,
\end{align}
where $\kappa_n=n\Omega=2\pi n/T\;, (n=0,\pm 1,\pm2, \cdots)$,
satisfying
\begin{align}
\hat p_t|\kappa_n)=\kappa_n|\kappa_n) \;.
\end{align}
The time basis is given by the transformation of
\begin{align}
|t)=\sum_{n=-\infty}^\infty e^{-i\kappa_n t}|\kappa_n) \;.
\end{align}
Together with the basis of ${\cal S}$-space, the complete orthonormal basis set in the ${\cal F}$-space is formed by $\{ |j,t)\!\>, |j^*,t)\!\>\}$  or $\{ |j,\kappa_n)\!\>, | j^*,\kappa_n)\!\>\}$ $(j=a,b_k)$ in terms of the time- or Floquet-mode-representations, respectively, where $|\cdot,\cdot )\!\>$ denotes a vector in the $\cal F$-space.
In this paper, we have abbreviated as $n\equiv\kappa_n$.
These basis satisfy the complete-orthonormality in the ${\C F}$-space as
\begin{align}
{\C I}={1\over T} \sum_j \int_0^T dt |j,t)\!\>\<\!(\tilde j,t| \;, \; \<\!(\tilde j,t|j',t')\!\>=T\delta(t-t')\delta_{j,j'} \;,
\end{align}
in terms of $\{ |j,t)\!\>\}$, or  
\begin{align}
1=\sum_{n=-\infty}^\infty \sum_j  |j,n)\!\>\<\!(\tilde j,n| \;, \; \<\!(\tilde j,n|j',n')\!\>=\delta_{n,n'}\delta_{j,j'} \;,
\end{align}
 in terms of $\{ |j,n)\!\>\}$ basis set.

Using the transform of (\ref{AppEq:kappan}),  the Floquet-Liouvillian ${\C L}_{\rm F}(t)$ (\ref{FloquetEq}) is represented in terms of the Floquet-mode representaion as
\begin{align}\label{AppEq:LFfull}
{\cal L}_{\rm F}=
\begin{array}{c| c c c c c  c c c c}
 &  |a,1)\!\> & |b_k,1)\!\>  & |a,0)\!\>  & |b_k,0)\!\> & \cdots & |a^*,1)\!\> & |b_k^*,1)\!\>  & |a^*,0)\!\> & |b_k^*,0)\!\>  \\ \hline
\<\!( \tilde a,1| &  -\omega_0-\Omega & -g_k & i f_0 & 0 & \cdots & 0   &  0  & if_0 &0\\
\<\!(\tilde b_k,1| &  -g_k & -\omega_k-\Omega & 0 & 0  & \cdots&  0   &    0  & 0  & 0\\
\<\!(\tilde a,0| & -i f_0 & 0 &-\omega_0  & -g_k & \cdots&  -i f_0  &0  &0  &0  \\
\<\!(\tilde b_k,0| & 0 &  0 &  -g_k  & -\omega_k & \cdots & 0 & 0 & 0 &0 \\ 
\vdots & \vdots & \vdots  & \vdots & \vdots & \vdots & \vdots & \vdots & \vdots & \vdots\\
 \<\!( \tilde a^*,1| &  0     & 0    & -if_0  & 0  & \cdots & \omega_0-\Omega  & g_k  & -if_0 &0  \\
  \<\!(\tilde b_k^*,1| & 0      & 0    & 0&  0 &  \cdots & g_k &   \omega_k-\Omega &0 & 0  \\ 
 \<\!( \tilde a^*,0| &  if_0 &0  & 0 & 0 & \cdots & if_0  &0  & \omega_0  & g_k  \\
 \<\!(\tilde  b_k^*,0| & 0 &0  &0  &0 & \cdots  &0  & 0  & g_k  & \omega_k  
\end{array}
\end{align}
where we show the matrix only for the $n=0$ and $n=1$ Floquet modes and a particular $k$ mode of the photonic band, for simplicity.
The complex eigenvalue problem of the Floquet-Liouvillian reads in time-independent form as given by
\begin{align}\label{AppEq:LF}
{\cal L}_{\rm F}|\varphi_\xi)\!\>=z_\xi|\varphi_\xi)\!\> \;,\;{\cal L}_{\rm F}|\B\varphi_\xi)\!\>=\B z_\xi|\B\varphi_\xi)\!\> \;.
\end{align}
It should be noted that while the virtual transition couplings make ${\cal L}_{\rm F}$ non-Hermitian, for example as
$
\<\!(\tilde a^*,1|{\cal L}_{\rm F}|a,0 )\!\>=\<\! (\tilde a,0|{\cal L}_{\rm F}|a^*,1 )\!\>=-i f_0
$,
${\C L}_{\rm F}$ holds the symplectic symmetry for an entire states
\begin{align}\label{AppEq:LFSymplec}
{\C J} {\C L}_{\rm F}{\C J}={\C L}_{\rm F}^T \;.
\end{align}
which ensures that the derived results are consistent with the microscopic dynamics.

Under the condition of 
\begin{align}
\Omega-2B \gg f_0 \;,
\end{align}
which is satisfied for the narrow photonic band and small external amplitude, we can restrict ourselves to  the $\{|a,0)\!\>, |b_k,0)\!\>, |a^*,1)\!\>, |b_k^*,1)\!\> \}$-subspace.

In order to maintain the symplectic symmetry, we shift the energy origin by $\Omega/2$ so that
\begin{align}\label{AppEq:LF0}
{\cal L}_{\rm F}^{(0)}=
\begin{array}{c| c c c c c c c c}
 &  |\tilde a,0)\!\>  & |b_k,0)\!\>  & |a^*,1)\!\> & |b_k^*,1)\!\>    \\ \hline
\<\!( \tilde a,0| &-\omega_0+{\Omega\over 2}  & -g_k & -i f_0  &0    \\
\<\!(\tilde  b_k,0|  &  -g_k  & -\omega_k+{\Omega\over 2}  & 0 & 0  \\ 
 \<\!( \tilde a^*,1|    & -if_0  & 0 & \omega_0-{\Omega\over 2}  & g_k    \\
  \<\!( \tilde b_k^*,1|     & 0&  0&   g_k &   \omega_k-{\Omega\over 2}  
\end{array}  \;,
\end{align}
where $|b_k,0)\!\>$ and $|b_k,1)\!\>$ state represents other continuous states for the $n=0$ and $n=1$ Floquet modes, respectively.
The shifted eigenvalue problems reads
\begin{align}\label{AppEq:EVLF0}
{\cal L}_{\rm F}^{(0)}|\varphi_\xi)\!\>=z'_\xi|\varphi_\xi)\!\> \;,\;{\cal L}_{\rm F}^{(0)}|\B\varphi_\xi)\!\>=\B z'_\xi|\B\varphi_\xi)\!\> \;.
\end{align}
where 
\begin{align}
z'_\xi\equiv z_\xi+{\Omega\over 2}
\end{align}

Using the projection operators \cite{Yamane18Symmetry} of 
\begin{align}
{\cal P}_{a}=   |a,0)\!\>\<\!(\tilde a,0|+ |a^*,1)\!\>\<\!( \tilde a^*,1|  \;,\; {\cal Q}_a\equiv 1-{\cal P}_a\;,
\end{align}
we have derived the effective Floquet-Liouvillian  \cite{Petrosky91Physica,Tanaka16PRA,Fukuta17PRA,Yamane18Symmetry,Tanaka18META}
\begin{align}
{\cal L}_{\rm eff}(z)\equiv {\cal P}{\cal L}_{\rm F} {\cal P}+ {\cal P}{\cal L}_{\rm F} {\cal Q}{1\over z'-{\cal Q}{\cal L}_{\rm F}{\cal Q}}  {\cal Q}{\cal L}_{\rm F} {\cal P} \;.
\end{align}
in the  $\{  |a,0)\!\>, |a^*,1)\!\> \}$-subspace as
\begin{align}\label{AppEq:Leff}
{\cal L}_{\rm eff}(z)=
\begin{array}{c|c c}
 &  |a,0)\!\> & |a^*,1)\!\>   \\ \hline
\<\!( \tilde a,0| &  {\Omega\over 2}-\omega_0-\sigma\left({\Omega\over 2}-z'\right) &  -if_0   \\
\<\!( \tilde a^*,1| &  - if_0   &-\left({\Omega\over 2}-\omega_0\right)+\sigma\left({\Omega\over 2}+z'\right) \\ 
\end{array} \;,
\end{align}
where in order to maintain the symplectic symmetry, we shift the energy origin by $\Omega/2$, yielding the effective Floquet-Liouvillian represented by a two-by-two matrix in terms of  $\{|a,0)\!\>, |b_k,0)\!\>, |a^*,1)\!\>, |b_k^*,1)\!\> \}$.
In (\ref{AppEq:Leff}) the self-energy is analytically obtained for the interaction of the cavity mode with the one-dimensional photonic crystal as
\begin{align}
\sigma(z)\equiv {B^2\over \pi}\int_{0}^\pi {g^2\sin^2k\over z-\omega_k} dk =g^2\left( z-\omega_B-\sqrt{(z-\omega_B)^2-B^2}\right) \;.
\end{align}

\section{Resonance eigenmode of  Floquet-Liouvillian}\label{AppSec:Modes}

In Section \ref{Sec:Spectra}, we have shown three different types of stationary modes, one stable mode and two resonance modes, by solving the complex eigenvalue problems of the effective Floquet-Liouvilia.
The difference of these stationary modes will be clarified when we observe the eigenmodes of the total system Floquet-Liouvilian.

Taking care of the analytic continuation, we consider the right-eigenstates corresponding to the resonance modes of the effective Liouvillian (\ref{Leff2}).
The complex eigenvalue problem of the effective Floquet-Liouvillian 
\begin{align} \label{LeffEV}
{\cal L}_{\rm eff}(z'_\xi){\C P}_a|\varphi_\xi)\!\>=z'_\xi{\C P}_a|\varphi_\xi)\!\>  \;,
\end{align}
 reads
\begin{align}
\begin{pmatrix}
  -\omega_0'+\sigma(z'+\omega_B') &  -if_0   \\
  - if_0   &  \omega_0'+\sigma(z'-\omega_B') \\ 
\end{pmatrix} 
\begin{pmatrix}
\<\!( \tilde a,0|\varphi_\xi)\!\>\\
\<\!( \tilde a^*,1|\varphi_\xi)\!\>
\end{pmatrix}
=z'_\xi
\begin{pmatrix}
\<\!( \tilde a,0|\varphi_\xi)\!\>\\
\<\!( \tilde a^*,1|\varphi_\xi)\!\>
\end{pmatrix} \;.
\end{align}

The ratio of the components for the canonical pair  of the creation mode $|\B\varphi_\xi)\!\> $  and the annihilation mode $|\varphi_\xi)\!\>$ are given by
\begin{align}\label{Ratio}
{ \<\!(\tilde a,0|\B \varphi_\xi)\!\> \over \<\!(\tilde a^*,1|\B\varphi_\xi)\!\>}
= { {\B z}_\xi'-\omega'_0-\sigma(\B z_\xi'-\omega'_B) \over -i f_0} \;, \;
{ \<\!(\tilde a^*,1|\varphi_\xi)\!\> \over \<\!(\tilde a,0|\varphi_\xi)\!\>}
= {-if_0\over  { z}_\xi'-\omega'_0-\sigma( z_\xi'-\omega'_B)} \;, 
\end{align}

The eigenstates of the total system is obtained by adding the component of the complementary space which is defined given by
\begin{align}
{\C Q}_a|\varphi_\xi)\!\>={1\over z-{\C Q}_a{\C L}_{\rm F}{\C Q}_a}{\C Q}_a{\C L}_{\rm F} {\C P}_a|\varphi_\xi)\!\> \;.
\end{align}
We then have
\begin{subequations}
\begin{align}
|\varphi_\xi)\!\>&=\<\!(\tilde a,0|\varphi_\xi)\!\>\left\{ |a,0)\!\>-\int dk {g_k\over z'_\xi+\omega'_B-\omega_k} |b_k,0)\!\> \right\}
+\<\!(\tilde a^*,1|\varphi_\xi)\!\>\left\{ |a^*,1)\!\>-\int dk {g_k\over z'_\xi-\omega'_B+\omega_k } |b_k^*,1)\!\> \right\}  \;,\\
 |\B\varphi_\xi)\!\>&=\<\!(\tilde a^*,1|\B\varphi_\xi)\!\>\left\{ |a^*,1)\!\>-\int dk {g_k\over \B z'_\xi-\omega'_B+\omega_k }|b_k^*,1)\!\> \right\}
 +\<\!(\tilde a,0|\B\varphi_\xi)\!\>\left\{ |a,0)\!\>-\int dk {g_k\over \B z'_\xi+\omega'_B-\omega_k }|b_k,0)\!\> \right\}  \;.
\end{align}
\end{subequations}
With the use of the relation of (\ref{RelationLR}), the left-eigenmode functions are similarly obtained as
\begin{subequations}
\begin{align}
\<\!( \tilde\varphi_\xi|&=\<\!( \tilde a^*,1|\B\varphi_\xi)\!\>\left\{ \<\!( \tilde a,0)|-\int dk {g_k\over \B z'_\xi -\omega'_B+\omega_k} \<\!(\tilde b_k,0| \right\}-\<\!(\tilde a,0|\B\varphi_\xi)\!\>\left\{ \<\!(\tilde a^*,1|+\int dk{g_k\over \B z'_\xi +\omega'_B-\omega_k}  \<\!(\tilde  b_k^*,1| \right\}  \;,\\
\<\!(\tilde{\B\varphi}_\xi |&=\<\!( \tilde a,0|\varphi_\xi)\!\>\left\{ \<\!(\tilde a^*,1|+\int dk {g_k\over  z'_\xi +\omega'_B-\omega_k}  \<\!(\tilde b_k^*,1| \right\}-\<\!(\tilde a^*,1|\varphi_\xi)\!\>\left\{ \<\!(\tilde a,0)|-\int dk {g_k\over  z'_\xi -\omega'_B+\omega_k} \<\!(\tilde b_k,0| \right\} \;.
\end{align}
\end{subequations}

The normalization constants $\<\!(\tilde a,0|\varphi_\xi)\!\>$ and $\<\!(\tilde a^*,1|\B\varphi_\xi)\!\>$ are determined by the normalization condition (\ref{BiComOrho}) for all the degrees of freedom including the photonic band states,  which reads
\begin{align}
1=\<\!(\tilde\varphi_\xi|a,0)\!\>\<\!(\tilde  a,0|\varphi_\xi)\!\>+\<\!(\tilde\varphi_\xi| a^*,1)\!\>\<\!( \tilde a^*,1|\varphi_\xi)\!\>
+\int dk \left\{ \<\!(\tilde\varphi_\xi|b_k,0)\!\>\<\!(\tilde  b_k,0|\varphi_\xi)\!\>+\<\!(\tilde\varphi_\xi|b_k^*,1)\!\>\<\!(\tilde  b_k^*,1|\varphi_\xi)\!\> \right\} \;.
\end{align}
With the use of (\ref{RelationLR})   we have
\begin{align}
\<\!(\tilde a^*,1|\B \varphi_\xi)\!\>\<\!(\tilde a,0|\varphi_\xi)\!\>=\left[1+{d\over dz}\sigma(\B z'_\xi-\omega'_B)
+{ \B z'_\xi-\omega'_0-\sigma(\B z'_\xi-\omega'_B ) \over \B z'_\xi+\omega'_0+\sigma(\B z'_\xi+\omega'_B ) }
\left\{ 1+{d\over dz}\sigma(z'+\omega'_B)\bigg|_{z'=\B z'_\xi} \right\}\right]^{-1} \;.
\end{align}

The eigenmode operators are obtained from these left-eigenmode functions with the use of the relations of   (\ref{EigenOp}) as a multimode Bogoliubov transform of
\begin{align}
\hat\varphi_\xi&=\<\!(\tilde a^*,1|\B\varphi_\xi)\!\>\left\{\hat a -\int dk {g_k\over \B z'_\xi -\omega'_B+\omega_k} \hat b_k \right\}
-\<\!(\tilde a,0|\B\varphi_\xi)\!\>\left\{ \hat a^\dagger+\int dk{g_k\over \B z'_\xi +\omega'_B-\omega_k} \hat b_k^\dagger \right\}  \;,\\
\hat{\B\varphi}_\xi&=\<\!(\tilde a,0|\varphi_\xi)\!\>\left\{ \hat a^\dagger +\int dk {g_k\over  z'_\xi +\omega'_B-\omega_k}  \hat b_k^\dagger \right\}
-\<\!(\tilde a^*,1|\varphi_\xi)\!\>\left\{\hat a-\int dk {g_k\over  z'_\xi -\omega'_B+\omega_k} \hat b_k\right\} \;.
\end{align}


\end{document}